\documentclass[usenatbib]{mn2e}

\def\be{\begin{equation}}
\def\ee{\end{equation}}
\def\kms{{\rm\,km\,s^{-1}}}
\def\kpc{{\rm\,kpc}}

\def\pc{{\rm\,pc}}

\def\Gyr{{\rm\,Gyr}}
\def\half{{\textstyle{1\over2}}}

\def\bfv{{\bf v}}
\def\bfx{{\bf x}}

\def\bfz{{\bf z}}
\def\bfA{{\bf A}}
\def\bfC{{\bf C}}

\def\bfQ{{\bf Q}}
\def\bfI{{\bf I}}

\def\df{{\caps df}}
\def\dfs{{\caps df}s}
\def\rms{{\caps rms}}
\newcommand{\avr}{{\caps avr}}
\newcommand{\avrs}{{\caps avr}s}

\newfont{\caps}{cmcsc10}
\newcommand{\gmc}{{\caps gmc}}
\newcommand{\gmcs}{{\caps gmc}s}
\def\r#1{R$_{#1}$}

\title[Velocity distribution in solar neighbourhood]
{The stellar velocity distribution in the solar neighbourhood}

\author[R. De Simone, X. Wu \& S.\ Tremaine]
{Richard S. De Simone, Xiaoan Wu\thanks{To whom correspondence should be
addressed.} and Scott Tremaine \\
Princeton University Observatory, Peyton Hall, \\
Princeton, NJ 08544-1001, USA;
e-mail: desimone\_r@yahoo.com, xawn@astro.princeton.edu,
tremaine@astro.princeton.edu}

\begin{document}

\maketitle

\begin{abstract}

\noindent
We explore the heating of the velocity distribution in the solar neighbourhood
by stochastic spiral waves. Our investigation is based on direct numerical
integration of initially circular test-particle orbits in the sheared
sheet. We confirm the conclusion of other investigators that heating by spiral
structure can explain the principal features of the age-velocity dispersion
relation and other parameters of the velocity distribution in the solar
neighbourhood.  In addition, we find that heating by strong transient spirals
can naturally explain the presence of small-scale structure in the velocity
distribution (``moving groups''). Heating by spiral structure also explains
why the stars in a single velocity-space moving group have a wide range of
ages, a result which is difficult to understand in the traditional model that
these structures result from inhomogeneous star formation. Thus we suggest
that old moving groups arise from irregularities in the Galactic potential
rather than irregularities in the star-formation rate.
\end{abstract}


\section{Introduction}

\label{sec:intro}

The velocity distribution function (\df) of stars in the solar neighbourhood
provides unique insights into the Galactic potential field, the dynamical
history of the Galactic disk, and the relationships between kinematics, age,
and metallicity for disk stars.

Let us define the Local Standard of Rest (LSR) to be a fictitious point that
coincides with the Sun at the present instant and travels in a circular orbit
around the Galactic centre. We introduce a rotating Cartesian coordinate
system with origin at the LSR, $x$-axis pointing
radially outward, $y$-axis pointing in the direction of Galactic rotation, and
$z$-axis pointing to the south Galactic pole. The \df\ $f(\bfv)$ is defined so
that $f(\bfv)d\bfv$ is the number of stars per unit volume with velocities in
the range $[\bfv,\bfv+d\bfv]$.  The standard empirical model for the past
century has been the \citet{Sch07} \df, which is a triaxial Gaussian of
the form
\be
f(\bfv) \propto \exp\bigg[-\half\sum_{i,j=1}^3\alpha_{ij}(v_i-\overline
v_i)(v_j-\overline v_j)\bigg],
\label{eq:sss}
\ee
where $\bfv\equiv (v_x,v_y,v_z)$. The two lowest moments of the \df\ are the
mean velocity, 
$\overline \bfv=(\overline v_x,\overline v_y,\overline v_z)$, and the
velocity-dispersion tensor, 
\be
\sigma^2_{ij}\equiv \overline{(v_i-\overline v_i) (v_j-\overline v_j)};
\ee
where $\overline X\equiv \int f(\bfv)X(\bfv)d\bfv/\int f(\bfv)d\bfv$, 
and the matrix $\sigma^2_{ij}$ is the inverse of the matrix $\alpha_{ij}$.

In a steady-state, axisymmetric galaxy (i) the mean velocity relative to the
LSR is tangential, so $\overline v_x=\overline v_z=0$; (ii) the tensors
$\alpha_{ij}$ and $\sigma_{ij}^2$ are both diagonal; (iii) the ratio
$\sigma_{xx}^2/\sigma^2_{yy}$ is determined by the local gravitational force
and its radial gradient (e.g. \citealt{Cha42}, \citealt{BT}; see also
eq.\ \ref{eq:axis-ratio} below).

The velocity dispersion of stars in the solar neighbourhood increases with age,
probably because the disk is ``heated'' by one or more dynamical mechanisms.
However, the interpretation of the observed age-velocity dispersion relation
(\avr) is uncertain: (i) One school models the \avr\ as a smooth power law,
$\sigma_{xx}(t)\propto t^p$, and interprets this behavior as evidence for a
continuous heating mechanism. Estimates of the exponent $p$ in the literature
span the range 0.2--0.5.  (ii) Some authors argue that the dispersion rises
steeply with age for stars $\la 5\Gyr$ old, and thereafter is relatively flat
\citep{Carl85,Gom97}, perhaps because the continuous heating mechanism
saturates once the dispersion is large enough. (iii) A third model is that the
dispersion does not increase smoothly with age. For example, \citet{fre91} and
\citet{qg00} argue that there are three discrete age groups ($t\la 3\Gyr$,
$3\Gyr\la t\la 10\Gyr$, $t\ga 10\Gyr$) with different dispersions, but within
each group there is no evidence for a correlation between dispersion and
age. Such groups might arise because the continuous heating saturates after
only 3 Gyr, and the higher dispersion of the oldest stars is due to a discrete
event such as a merger.

A wide variety of mechanisms for disk heating has been discussed (see
\citealt{lac91} for a review): (i) Spitzer \& Schwarzschild (1951, 1953)
suggested that massive gas clouds could gravitationally scatter stars, leading
to a steady increase in velocity dispersion with age, and thereby predicted
the existence of giant molecular clouds (\gmcs). However, heating by \gmcs\
alone fails to explain several observations
\citep{lac84,vil85,jb90,lac91,jen92}: the predicted ratio
$\sigma_{zz}/\sigma_{xx}$ of vertical to radial dispersion may be too high,
roughly 0.72 compared to the observed value of 0.5 for old stars (but see
\citealt{ikm93} for an opposing view); the
predicted exponent in the \avr\ is somewhat too low, $p\la 0.25$; and the
masses and number density of \gmcs\ determined from CO observations are too
low to explain the observed dispersion, probably by a factor of five or so but
with substantial uncertainty.  (ii) Transient spiral waves lead to potential
fluctuations in the disk that excite the random motions of disk stars
\citep{bw67,sc84,cs85}. However, spiral waves only excite the horizontal ($x$
and $y$) velocity components effectively, since their characteristic spatial
and temporal scales are much larger than the amplitude or period of vertical
oscillations for young stellar populations. (iii) These considerations lead
naturally to a hybrid model, in which spiral waves excite non-circular
velocities in the plane, and the velocities are then redistributed between
horizontal and vertical motion through \gmc\ scattering
\citep{car87}. The hybrid model has been investigated thoroughly by
\citet{jb90} and \citet{jen92} using the Fokker-Planck equation. They
find that they can successfully reproduce the observed axis ratio
$\sigma_{zz}/\sigma_{xx}$, the exponent $p$ in the \avr\ and the
radial dispersion of old stars.  (iv) Other possible heating
mechanisms, all of which rely to some extent on hypothetical or poorly
understood components or features of the Galaxy, include scattering by massive
compact halo objects or halo substructure \citep{lo85}, mergers with dwarf
galaxies \citep{to92,wmh96,hc97}, or the outer Lindblad resonance from the
Galactic bar \citep{kal91,deh99,deh00,fux01,qui03}.

The Schwarzschild \df\ (\ref{eq:sss}) only approximates the velocity
distribution on the largest scales in velocity space. On smaller scales, there
is substructure, which is most prominent in the youngest stars but present in
stars of all ages. Discussion of substructure in the velocity \df\ dates back
to Kapteyn's (1905) model of ``two star streams'', and for decades Eggen
advocated the case for substructure in the form of ``moving groups'' in the
solar neighbourhood (\citealt{egg96} and references therein). 

Eggen and others have usually explained moving groups as the result of
inhomogeneous star formation in the disk: in this model, stars in a moving
group are born at a common place and time, and then disperse into a stream
that happens to intersect the solar neighbourhood. This model predicts that
stars in a moving group should have the same age, metallicity, and azimuthal
velocity (i.e. the same angular momentum, since this determines their mean
angular velocity). The existence and membership of these groups was
controversial until the Hipparcos satellite measured reliable distances and
proper motions for a large, homogeneous database of nearby stars, and verified
the presence of rich substructure in the velocity \df\ of both young and old
stars, including a number of features that coincide with moving groups already
identified by Eggen (e.g. \citealt{deh98}, Chereul et al.\
1998,1999)

In this paper we explore a quite different explanation for substructure in the
velocity \df. We suggest that substructure arises naturally from the same
spiral gravitational fluctuations that excite the growth of the velocity
dispersion. In other words, substructure is caused by homogeneous star
formation in an irregular potential, as opposed to inhomogeneous star
formation in a regular potential in the traditional model.

We investigate this hypothesis by simulating the evolution of the velocity \df\
induced by transient spiral structure in a simple two-dimensional model of the 
local Galaxy. We restrict ourselves to two dimensions because spiral structure 
does not excite vertical motions efficiently, and because the velocity \df\ 
appears to be well-mixed in the vertical direction \citep{deh98}. 

Following Eggen, we shall use the term ``moving group'' to denote substructure
in the velocity \df\ of old stars ($\ga 1$ Gyr) at a given
position. Unfortunately, the same term is sometimes also applied to OB
associations, which are spatially localized concentrations of much younger
stars (e.g., \citealt{dez99}).

Section \ref{sec:MODEL} describes our simplified dynamical model. The results
of our simulations are analyzed and compared to observed data in
\S\ref{sec:RESULTS}. Section \ref{sec:inhomo} examines briefly the traditional
hypothesis that substructure arises from inhomogeneous star formation. Section
\ref{sec:ORBIT} contains a brief discussion of the closely related process of
radial migration of stars, and \S\ref{sec:CONCLUDE} contains concluding
remarks.

\section{Properties of spiral structure}

\label{sec:prop}

Our model depends on several properties of the Galaxy's spiral structure, such
as the number of arms, the arm strength, and the pitch angle. We are concerned
with spiral structure in the disk surface density (rather than, say, in the
distribution of young stars or gas). The properties of this structure in
external galaxies are revealed by near-infrared images, which are dominated by
the stars that contribute most of the mass (\citealt{rr93}, but see also
\citealt{rho98} and \citealt{js99} for qualifications).

\citet{rz95} examined the K-band spiral structure of 18 face-on
galaxies. They found that almost half had strong two-arm ($m=2$) spirals, with
arm-interarm contrasts $I_{\rm max}/I_{\rm min}\simeq 2$. If the arms are
sinusoidal, with fractional amplitude $\epsilon$ relative to the
axisymmetric background, the arm-interarm contrast is $I_{\rm max}/I_{\rm
min}=(1+\epsilon)/(1-\epsilon)$, so a contrast of 2
corresponds to $\epsilon\simeq 0.3$.

\citet{bp99} examined K-band spiral structure in 19 spirals. They found pitch
angles ranging from $8\degr$ to $49\degr$ with a median value of $22\degr$,
and $m=2$ amplitudes ranging from $\epsilon=0.03$ to $\epsilon=0.5$, with a median $\epsilon=0.1$.

\citet{sj98} conducted a similar survey of 45 galaxies. They found that the
dominant Fourier mode usually has $m=1$ (36\%) or $m=2$ (31\%), and the median
pitch angle of the spiral structure was $8\degr$, with little or no
correlation with Hubble type. They measure the strength of the spiral arms in
terms of the ``equivalent angle''; their median equivalent angle of $14\degr$
corresponds to an amplitude $\epsilon\simeq 0.07$ for a sinusoidal
$m=2$ spiral.

\citet{elm99} stress that the amplitude of $m=2$ spiral structure in the
near-infrared depends on whether the optical spiral arms are classified as
flocculent or grand-design \citep{elm98}. Grand-design spirals have
arm-interarm contrasts of 1.5--6, corresponding to amplitude
$\epsilon=0.2$--0.7, while flocculent galaxies have contrast $\la 1.7$,
corresponding to $\epsilon\la 0.25$.

At optical wavebands, \citet{ma99} find that the mean pitch angle for 51 Sbc
galaxies (the same Hubble type as the Galaxy) is $15\degr$.

The measurement of spiral structure in our own Galaxy is more difficult than
in face-on external galaxies. \citet{dri00} uses K-band photometry of the
Galactic plane to conclude that the Galaxy contains a two-arm spiral with
pitch angle $18\degr$. \citet{val02} reviews a number of studies of the
Galaxy's spiral structure, mostly based on young stars, gas and dust; these
yield a range of pitch angles from $6\degr$--$20\degr$, but Vall\'ee concludes
that the best overall fit is provided by an $m=4$ spiral with pitch angle of
$12\degr$. The arm structure in the Galaxy appears to be intermediate between
grand-design and flocculent \citep{elm98}.

\section{A numerical model of disk heating} 

\label {sec:MODEL}

We model disk heating in the sheared sheet, which approximates the local
dynamics of a differentially rotating disk \citep{SS53,glb65,jt66}. The LSR is
assumed to travel around the Galactic centre in a circular orbit of radius
$R_0$ and angular frequency $\Omega>0$. We use the same Cartesian coordinate
system $(x,y,z)$ introduced in the last Section, restrict ourselves to the
$z=0$ plane, and denote position and velocity by $\bfx=(x,y)$ and $\bfv=(\dot
x,\dot y)$. For $|x|,|y|\ll R_0$ the equations of motion of a test particle
are
\begin{eqnarray}
\ddot x - 2 \Omega \dot y - 4 \Omega A x & = & - 
\frac{\partial \Phi}{\partial x} , \nonumber\\
\ddot y + 2 \Omega \dot x & = & - \frac{\partial \Phi}{\partial y} \label
{eq:y-frame}, 
\end{eqnarray}
where $\Phi(\bfx,t)$ is the gravitational potential due to sources other than
the axisymmetric Galactic disk, and $A>0$ is the Oort constant
\be
A = - \frac{1}{2}\left(R \frac{d \Omega}{d R}\right)_{R_0},
\ee
where $R\Omega(R)$ is the circular speed at radius $R$ and
$\Omega(R_0)=\Omega$. 

If $\Phi$ = 0 the trajectories governed by equations (\ref{eq:y-frame}) have 
two integrals of motion related to energy and angular momentum,
\be
E\equiv \half(\dot x^2 + \dot y^2 -4\Omega A x^2), \qquad H\equiv \dot
y+2\Omega x.
\label{eq:ehh}
\ee
The solutions of the equations of motion are 
\begin{eqnarray}
x & = & x_g + a \cos(\kappa t + \phi), \nonumber\\ y & = & y_g(t) - \frac{ 2
\Omega}{\kappa} a \sin(\kappa t + \phi), \nonumber \\
y_g(t) & = & y_{g0} - 2Ax_g t,
\label{y-epi} 
\end{eqnarray}
where $[x_g,y_g(t)]$ are the coordinates of the guiding centre, $a$ is the
epicycle amplitude, $\phi$ is a phase constant, and $\kappa$ is the radial or
epicycle frequency,
\be
\kappa = \Omega \left( 4 + 2\frac{d \ln \Omega}{d \ln R}
\right)^{1/2}_{R_0}\quad \hbox{or}\quad \kappa^2=4\Omega(\Omega-A). 
\label{kappa-def} 
\ee
The energy and angular momentum are related to the guiding-centre radius and
epicycle amplitude by
\be
E=2A(A-\Omega)x_g^2+\half\kappa^2a^2, \qquad H=2(\Omega-A)x_g,
\label{eq:hxg}
\ee
and the guiding-centre radius is related to the phase-space coordinates by
\be
x_g={\dot y+2\Omega x\over 2(\Omega-A)}.
\ee

The epicycle energy is defined as
\begin{eqnarray} 
E_x & \equiv & \half\left[\dot x^2 + \kappa^2(x - x_g)^2\right] \nonumber \\
    & = & E+{AH^2\over 2(\Omega-A)} \nonumber \\
    & = & \half\kappa^2a^2 \nonumber \\
    & = & \half\dot x^2 +{2\Omega^2\over\kappa^2}(\dot y + 2Ax)^2. 
\label{eq:epicycle-energy}
\end{eqnarray}

The epicycle energy is closely related to the radial action
\be
I=\half\kappa a^2=E_x/\kappa.
\label{eq:action}
\ee
For a particle in a circular orbit, 
\be
(x,y,\dot x,\dot y)=(x_g,y_{g0}-2Ax_gt,0,-2Ax_g),
\ee
thus ${E_x=0}$.

The approximations used in deriving the linearized equations of motion
(\ref{eq:y-frame}) are only marginally valid in the solar neighbourhood, since
the epicycle amplitudes $a$ can be a significant fraction of $R_0$ (for a
population with radial velocity dispersion $\sigma_{xx}$ in a galaxy with a
flat rotation curve,
\be
\overline{a^2}=\sigma_{xx}^2/\Omega^2.
\label{eq:epiamp}
\ee
Thus old disk stars in the solar neighbourhood, with $\sigma_{xx}\simeq40\kms$,
have $(\overline{a^2})^{1/2}\simeq 0.18R_0$). Nevertheless, we believe that
the sheared sheet accurately captures the most important features of the
evolution of disk-star kinematics.

The kinematics of a population of stars is described by the \df\
$f(\bfx,\bfv,t)$, where $fd\bfx d\bfv$ is the number of stars in the interval
$(\bfx,\bfv)\to(\bfx+d\bfx,\bfv+d\bfv)$. According to Jeans's theorem
\citep{BT}, a stationary \df\ can only depend on the integrals of motion $E$
and $H$ (or $E_x$, $x_g$, $a$, etc.). In the solar neighbourhood ($x=y=0$),
the integrals are $H=\dot y$ and $E_x=\half\dot x^2+2\Omega^2\dot
y^2/\kappa^2$. Thus in a steady state the velocity distribution must be an
even function of $\dot x$.

It is straightforward to show that the mean velocity and velocity-dispersion
tensor of any stationary, spatially homogeneous \df\ in the sheared sheet must
satisfy the relations
\be
\overline v_x=0, \ \overline v_y=-2Ax, \  
\sigma^2_{xy}=\sigma^2_{yx}=0, \ 
{\sigma^2_{xx}\over\sigma^2_{yy}}={\Omega \over\Omega-A}.
\label{eq:axis-ratio}
\ee

A useful model \df\ for the sheared sheet is
\be
f(\bfx,\bfv)\propto \exp\left(-E_x/\sigma_{0}^2\right), 
\label{eq:Schwarz}
\ee
where $E_x$ is defined in equation (\ref{eq:epicycle-energy}). At a given
location, the \df\ (\ref{eq:Schwarz}) leads to a two-dimensional version of
the triaxial Gaussian velocity distribution (\ref{eq:sss}), in which the mean
velocity and velocity-dispersion tensor satisfy the relations
(\ref{eq:axis-ratio}). Thus the \df\ (\ref{eq:Schwarz}) is also sometimes
(confusingly) called the Schwarzschild \df.

In our simulations the relations (\ref{eq:axis-ratio}) are not satisfied
exactly, so it is useful to introduce the principal-axis system
$(x_1,x_2)$ in which the velocity-dispersion tensor is diagonal; the
$x_1$-axis is chosen to lie within $45\degr$ of the $x$-axis. The
corresponding velocity dispersions are $\sigma_1$ and $\sigma_2$, which we
will normally plot instead of $\sigma_{xx}$ and $\sigma_{yy}$; usually in our
simulations and in the data we shall find that $\sigma_1>\sigma_2$.  The
vertex deviation $l_v$ is the Galactic longitude of the $x_1$-axis, and is
given by
\be
l_v \equiv -\frac{1}{2} \arctan{\left(\frac{2\sigma^2_{xy}}{\sigma^2_{xx} -
\sigma^2_{yy}}\right)}, \label{eq:vert} 
\ee
where $|l_v|<45\degr$. Stellar populations in the solar neighbourhood have
vertex deviations in the range $0\la l_v\la 30\degr$ \citep{bm98}.

Throughout the paper, we assume that the rotational curve of the underlying
axisymmetric galaxy is flat, $R\Omega(R)=\hbox{constant}$, so that
$A={1\over2}\Omega$ and $\kappa=\sqrt{2}\Omega$.
We shall assume that the surface density of the disk in the solar
neighbourhood is $\Sigma_d=50M_\odot\pc^{-2}$; recent observational estimates
are $\Sigma_d=40M_\odot\pc^{-2}$ \citep{cre98}, $48M_\odot\pc^{-2}$
\citep{hf00}, and $42\pm6M_\odot\pc^{-2}$ \citep{kor03}. The assumed surface
density only enters our analysis through the definition of
the fractional spiral amplitude $\epsilon$ below (eq.\ \ref{eq:phisdef}).

\subsection{Spiral waves}
\label{sec:spiral}

We approximate spiral structure as a superposition of waves with surface
density and potential 
\begin{eqnarray}
\Sigma (x,y,t) & = & \Sigma_{s}(t)\, \exp i(k_x x + k_yy),\nonumber  \\ 
\Phi (x,y,t) & = & \Phi_{s}(t)\, \exp i(k_x x + k_y y),\label{spirals-pot}
\end{eqnarray}
where ${\bf k}=(k_x,k_y)$ is the wavenumber, and only the real part of
$\Sigma$ or $\Phi$ is physical. Without loss of generality we can
assume that $k_y\ge 0$; then the spiral is trailing if $k_x>0$ and leading if
$k_x<0$. In this paper, we only consider trailing spiral waves. 
The number of arms $m$ and the pitch angle $\alpha$ are determined by
${\bf k}$ through the relations 
\be
k_y = \frac{m}{R_0} , \hspace{.5in} \left|\frac{k_y}{k_x}\right| = \tan\alpha.
\label{wave-number}
\ee
The relation between potential and surface density for spiral waves is given
by (e.g. \citealt{BT})
\be
\Phi_{s}(t) = - \frac{2 \pi G \Sigma_{s}(t)}{k}, 
\label{den-pot}
\ee
where $k=|{\bf k}|$. 

Steady spirals, in which $\Sigma_s(t),\Phi_s(t)\propto \exp(i\omega t)$, heat
stars on nearly circular orbits only at the Lindblad resonances
\citep{Lynd1972}, which occur at the guiding centre radii given by
\be
\omega=2Ak_yx_g\pm\kappa.
\label{eq:lind}
\ee
We focus instead on transient spiral structure, which can heat
stars over a range of radii.  We model each transient spiral using a
Gaussian amplitude dependence centred at time $t_s$ with standard deviation
$\sigma_s$: 
\be
\Phi_s(t)={2\pi\epsilon G\Sigma_d\over k}\exp \left[-\frac{(t-t_s)^2}{2
\sigma_s^2} + i(\theta+2Ak_yx_ct)\right].
\label{eq:phisdef}
\ee
Here $\theta$ is a phase constant, and $x_c$ is the corotation radius relative
to the LSR (the phase of the spiral is constant for an observer on a circular
orbit at $x_c$). The parameter $\epsilon$ measures the amplitude of the
spiral, and the normalizing constants are chosen so that the maximum surface
density in the spiral is a fraction $\epsilon$ of the surface density of the
underlying disk (eq.\ \ref{den-pot}). 

In each simulation the trajectories of the stars were followed between time
$t=0$ and $t=t_0$. During this interval the stars were perturbed by $N_s$
transient spiral waves. Each wave had phase constant $\theta$ chosen randomly
from $[0,2\pi]$, corotation radius $x_c$ chosen randomly from a Gaussian
distribution centred on the LSR with standard deviation $\sigma_c$, and
central time $t_s$ chosen randomly from $[-2\sigma_s, t_0+2\sigma_s]$ (this is
slightly longer than the integration interval, to include the effects of
transients whose wings are inside the integration interval although their
central times are not).

The fractional amplitude of the surface-density perturbation due to a single
transient spiral is $\epsilon$ at the wave peak. However, in some ways a
better quantity to compare with the observational data on spiral amplitudes is
the root-mean-square (\rms) time average of the fractional surface-density
amplitude,
\be
\epsilon_{\rms}=\epsilon\left(\pi^{1/2}N_s\sigma_s\over t_0\right)^{1/2}.
\label{eq:epsrms}
\ee
A closely related quantity is the \rms\ potential perturbation, 
\be
\Phi_{\rms}=2\pi G\epsilon_{\rms}\Sigma_dR_0{|\sin\alpha|\over \sqrt{2}m};
\label{eq:phirms}
\ee
(the factor $\sqrt{2}$ arises because $\Phi_{\rms}$ is the \rms\ fluctuation
rather than the \rms\ amplitude, which is smaller by $\sqrt{2}$). 
\citet{jb90} use Fokker-Planck calculations of heating by transient spiral
structure to estimate that $\Phi_{\rms}=(9\hbox{--}13\kms)^2$.

The power spectrum of each transient is a Gaussian with standard deviation
$(2\sigma_s^2)^{-1/2}$. The central frequencies of the power spectra of the
transients follow a Gaussian distribution with standard deviation
$2Ak_y\sigma_c$. Thus the power spectrum of the ensemble of $N_s$ transients
is smooth if the overlap factor 
\be
C\equiv N_s{(2\sigma_s^2)^{-1/2}\over 2Ak_y\sigma_c}={N_s\over
2^{3/2}Ak_y\sigma_c\sigma_s} 
\label{eq:overlap}
\ee 
is large compared to unity; on the other hand if $C\ll1$ the power spectrum
and hence the heating is localized at narrow resonances. All of our
simulations have $C\gg1$ (see Table \ref{phys-param}). 

The dispersion of the power spectrum of all the transients is $\sigma_t$,
given by
\be
\sigma_t^2={1\over2\sigma_s^2}+(2Ak_y\sigma_c)^2.
\label{eq:st}
\ee

Note that in this model problem different azimuthal wavenumbers $m$ and
$m'=fm$ are equivalent if we rescale the other variables by: ${\bf k}'=f{\bf
k}$, $\alpha'=\alpha$, $t'=t$, $(x',y')=(x/f,y/f)$, $\Phi_s'=\Phi_s/f^2$,
$\Sigma_s'=\Sigma_s/f$, $\epsilon'=\epsilon/f$, $\sigma'_{ij}=\sigma_{ij}/f$
$\sigma_c'=\sigma_c/f$, $\sigma_0'=\sigma_0/f$, $\sigma_s'=\sigma_s$.

\subsection{Determining the solar neighbourhood velocity
distribution\label{sec:integration}} 

\label{sec:DIST}

Assume that the disk stars are formed on circular orbits at time $0$. We wish
to determine the velocity distribution at the present time $t_0$ in the solar
neighbourhood, $(x,y)=(0,0)$. This is a two-point boundary-value problem
rather than an initial-value problem: the boundary conditions on the
phase-space coordinates are $(x,y,\dot x,\dot y)=(x_g,y_{g0},0,-2Ax_g)$
 at $t=0$ (on circular
orbits) and $(x,y)=(0,0)$ at $t=t_0$ (in the solar neighbourhood).  The
solutions to the boundary-value problem will be a set of distinct initial
positions $[x_i(t=0),y_i(t=0)]$ or final velocities $[\dot x_i(t=t_0),\dot
y_i(t=t_0)]\equiv (-u_i,v_i)$, $i=1,2,\ldots$ (the sign of $u$ is chosen to
agree with the usual convention that stars moving towards the Galactic centre
have $u>0$). Because the motion of stars in the fluctuating gravitational
field of the transient spirals is complicated, we expect that there will be
many solutions. In this idealized model the velocity-space \df\ in the solar
neighbourhood will consist of a set of delta-functions at the velocities
$(u_i,v_i)$, but in practice these will be smeared into a continuous
distribution by observational errors, the non-zero volume of the ``solar
neighbourhood'', the small initial velocity dispersion of the stars when they
are formed, etc.

Normally, boundary-value problems are solved most efficiently by iterative
methods. In this case, however, most of the solutions have only a tiny domain
of attraction in the $(u,v)$ plane; thus iterative methods are inefficient. We
have therefore adopted a different approach, based on Monte Carlo sampling
\citep{deh00}.

Stars are not born on precisely circular orbits, in part because star-forming
clouds are not on circular orbits. We may therefore assume that the stars
initially have a Schwarzschild \df\ (eq.\ \ref{eq:Schwarz}), with a small but
non-zero initial velocity dispersion $\sigma_{0}$.

Following Dehnen, we integrate orbits backward in time, starting at $t=t_0$
(which we call the initial integration time or ``present epoch'') and ending
at $t=t_0$ (the final integration time or ``formation epoch''). The initial
conditions are chosen at random from $(x,y,\dot x,\dot y)=(0,0,-u,v)$, with
$|u|,|v|<v_{\rm max}$. We choose $v_{\rm max}=110\kms$, large enough to
include almost all disk stars in the solar neighbourhood.  We then integrate
equations (\ref{eq:y-frame}) backwards in time to the formation epoch. The
collisionless Boltzmann equation \citep{BT} states that the phase-space
density around a trajectory is time-invariant. Therefore the phase-space
density $F_i$ at $(u_i,v_i)$ at the {\em present} epoch is given by
(\ref{eq:Schwarz}), where the epicycle energy $E_x$ is measured at the {\em
formation} epoch. We assume that the spatial volume of our survey of the
solar neighbourhood is independent of velocity, so the density in velocity
space at the present epoch is proportional to $F_i$. Dehnen's procedure
therefore provides a Monte Carlo sampling of the velocity-space \df\ at the
present epoch.

We convert our Monte Carlo realization to a smooth \df\ by convolution with
observational errors. We replace the point solutions $\bfv_i=(u_i,v_i)$ by
Gaussians, that is,
\be
f(\bfv)=\sum_i F_i\delta(\bfv-\bfv_i)
\ee
is replaced by
\be
f(\bfv)={1\over2\pi\sigma_{\rm ob}^2} \sum_i
F_i\exp\left[-{(\bfv-\bfv_i)^2\over 2\sigma_{\rm ob}^2}\right],
\label{eq:error}
\ee
where $\sigma_{\rm ob}$ is the assumed observational error.

We have described the procedure for estimating the \df\ of stars of age $t_0$,
using equation (\ref{eq:Schwarz}) and the epicycle energies at $t=0$. However,
we can also determine the \df\ for stars having an intermediate age $t_0-t_m$
in the course of the same orbit integration, using the epicycle energies of
the stars at the intermediate time $t_m$. This approach measures the
dependence of the \df\ on stellar age, and thus determines the \avr\ for a
given realization of the set of transient spiral arms, using only a single
set of orbit integrations.

\subsection{Units and model parameters}

\label{sec:units}

The time unit is chosen to be $\Omega^{-1}$ and the distance unit is chosen to
be $R_0$. The azimuthal orbital period is then $2\pi/\Omega=2\pi$. For
conversion to physical units we assume that $R_0=8\kpc$ and $\Omega
R_0=220\kms$, so that the unit of time is 35.6 Myr. Most of our numerical
integrations ran for an interval $t_0=281$, corresponding to 40.8 rotation
periods of the LSR or 10.0 Gyr.

We consider only trailing spirals with $m=2$ or $m=4$. The standard deviation
of the distribution of corotation radii for the transient spirals was
$\sigma_c=0.25$ corresponding to 2 kpc.  With these choices the overlap factor
is $C=1.41N_s(m/2)/(\Omega\sigma_s)$.

The radial velocity dispersion of stars at the time of their birth is taken to
be $\sigma_0=3\kms$.  The observational error in the velocities is taken to be
$\sigma_{\rm ob}=3\kms$; for the typical Hipparcos parallax error of 1 mas,
$\sigma_{\rm ob}$ corresponds to the velocity error arising from the distance
uncertainty for a star with transverse velocity $30\kms$ at a distance of 100
pc.

\begin{figure*}
\begin{center}
\vspace{15cm}
\includegraphics{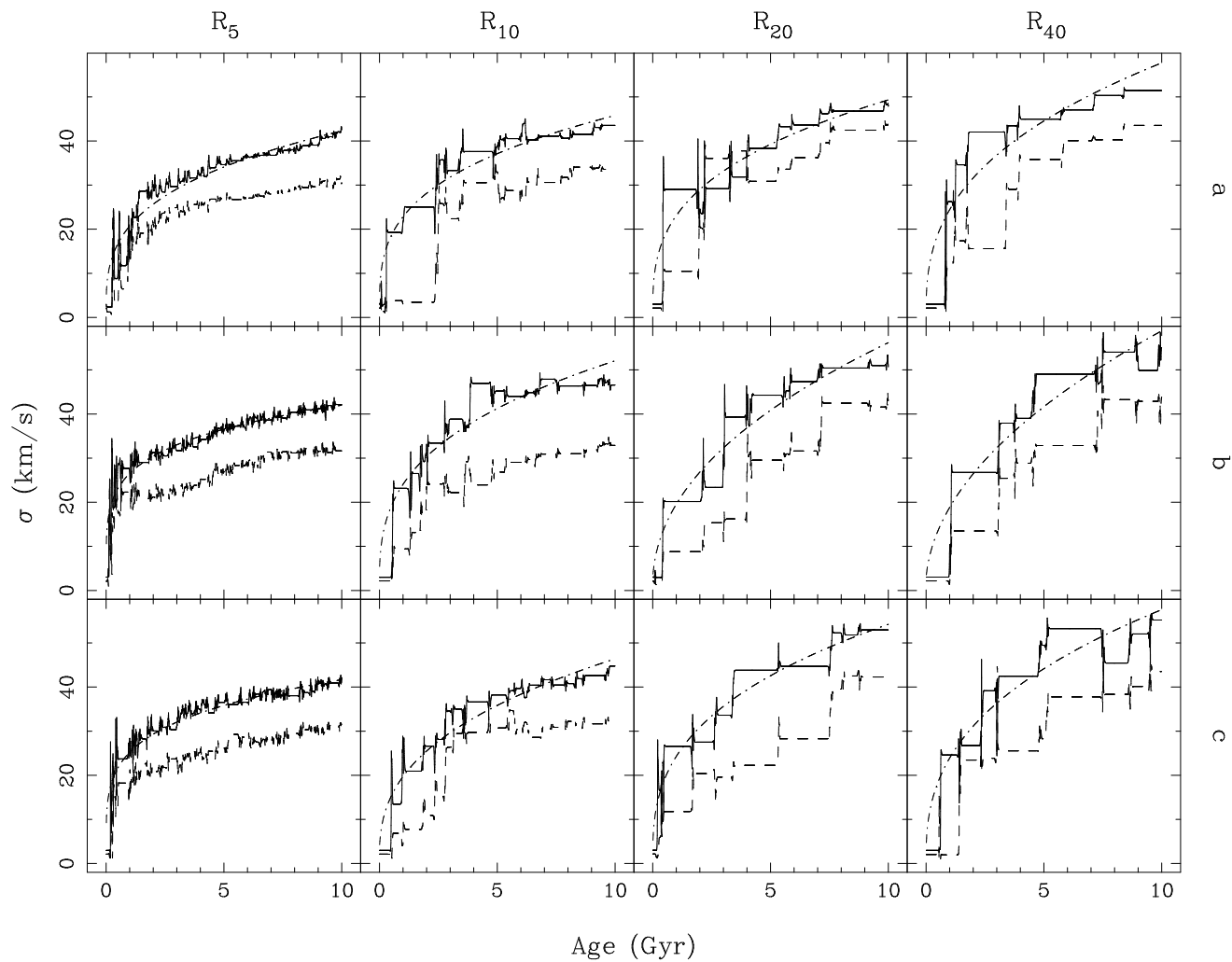}
\caption{The age dependence of the principal axes of the velocity-dispersion
tensor in runs \r5, \r{10}, \r{20} and \r{40}. The solid line shows $\sigma_1$
and the dashed line shows $\sigma_2$. The three rows (a,b,c) represent
different realizations. The dot-dashed lines show the best fit of equation
(\ref{eq:hist-fit}) for each realization; see also Table \ref{p-table}.}
\label{fig:age_disp}
\end{center}
\end{figure*}

In each simulation we launch either $1\times 10^7$ or $7\times10^7$ stars from
the solar neighbourhood and integrate back for 10 Gyr (the smaller simulations
are used to estimate moments, such as those in Figure \ref{fig:age_disp}, and
the larger simulations are used to estimate the \df, as in Figure
\ref{fig:contour}). In the larger simulations, typically 10000--15000 stars
contribute significantly to the solution, in the sense that their epicycle
energy at the formation epoch is $<3\sigma_0^2$. This number is larger than
the number of stars in the Hipparcos samples to which we compare (e.g.\ 4600
in the largest color-selected sample analysed by
\citealt{deh98}). We have verified that the conclusions below are unaffected
if we double the number of stars in the simulation or change the random-number
seed used to select the sample of stars (keeping the same seed to select the
spiral transients). 

\begin{table*}
\begin{minipage}{16cm}
\caption{Simulation parameters}
\begin{tabular}{|lccccccc|} 
Parameter & Symbol               & Run \r5 & Run \r{10} & Run \r{20} & Run
\r{40} & Run \r{m4} & Run \r{sim} \\ \hline 
\multicolumn{8}{c}{Defining parameters} \\
Number of spiral arms  &  $m$    & 2   &2  &2    &2   &4   &2  \\
Number of spiral waves & $N_{s}$ & 45 & 18 & 12  & 9 & 12 & 48 \\
Maximum fractional surface &     &       &       &       \\
density of spiral waves & $\epsilon$ & 1.61  & 0.80 & 0.80 &  0.67 & 0.80 & 0.31 \\
Time scale of spiral waves & $\sigma_s$ & 1 & 1 & 1 & 1 & 1 & 1\\
\rms\ corotation radius $(R-R_0)/R_0$ & $\sigma_c$ & 0.25 & 0.25 & 0.25 & 0.25 &0.25 &0.25 \\
Pitch angle          & $\alpha $ & $5\degr$ & $10\degr$ & $20\degr$ & $40\degr$ &$20\degr$ & $20\degr$ \\
\multicolumn{8}{c}{Derived parameters} \\
Overlap factor (eq. \ref{eq:overlap}) & $C$ & 64 & 25 & 17 & 13 & 34 & 68  \\ 
\rms\ surface-density amplitude (eq.~\ref{eq:epsrms}) & $\epsilon_{\rms}$ &
0.86 & 0.27 & 0.22 & 0.16 & 0.22 & 0.17 \\
\rms\ potential in $(\kms)^2$ (eq.~\ref{eq:phirms}) & $\Phi_{\rms}$ &
$(16.9)^2$ & $(13.4)^2$ & $(17.0)^2$ & $(19.8)^2$ & $(12.0)^2$ & $(15.0)^2$ \\
\hline
\end{tabular}
\label{phys-param} 
\end{minipage}
\end{table*}

We carried out a number of simulations with different values of the
spiral-wave parameters $m$ (number of arms), $N_s$ (number of spirals),
$\epsilon$ (fractional surface-density perturbation), $\sigma_s$ (duration of
the transient), $\sigma_c$ (dispersion in corotation radius), and $\alpha$
(pitch angle). The parameters of the simulations are shown in Table
\ref{phys-param}. Each simulation is defined by these six parameters  and the
random number seeds for both the stars and the spiral arms.

Our runs have pitch angles of $5\degr$, $10\degr$, $20\degr$, and $40\degr$,
which span the range of observed pitch angles in spiral galaxies. We then
choose the amplitude $\epsilon$ and the number of transients $N_s$ to
approximately reproduce the observed radial velocity dispersion of old
main-sequence stars in the solar neighbourhood\footnote{More precisely,
main-sequence stars redward of the Parenago discontinuity at $B-V=0.6$ are all
believed to have the same mean age, and have a color-independent radial
dispersion $\simeq 38\kms$
\citep{DehBin1998}. We match this to the radial dispersion in our simulation
by assuming a constant star-formation rate, which is consistent with solar
neighbourhood data within the large uncertainties \citep{Bin00}.}.

All of our runs have two-armed spirals ($m=2$), except for run \r{m4}, which
has the same parameters as run \r{20} except that $m=4$.

We also have run \r{sim}, in which we have used a larger number of weaker
transients than in run \r{20}, and have also adjusted some of the central
times of the spiral transients to better reproduce the features of the
observed solar neighbourhood \df\ (see \S\ref{sec:obsdf}). 

The \rms\ potential fluctuation $\Phi_{\rms}=(13\hbox{--}20\kms)^2$ in our
four runs, higher than the velue $\Phi_{\rms}=(9\hbox{--}13\kms)^2$ derived by
\citet{jb90} from Fokker-Planck calculations by about a factor of 2--2.4. A
minor component of this difference arises because \citet{jb90} include
\gmcs, which contribute about 20\% of their total heating rate. A more
important effect, pointed out to us by A.~J.~Kalnajs, is that in our model the
heating is spatially inhomogeneous: our assumptions place the Sun fairly close
to corotation for most of the transients, so that heating at the Lindblad
resonances is more effective inside and outside the solar circle than it is at
the solar circle.

The fractional surface-density fluctuation $\epsilon$ required to reproduce
the \avr\ in run \r{5} exceeds unity, and the \rms\ surface-density
fluctuation is close to unity; these high values are unrealistic and suggest
that a larger number of weaker transients is required if the pich angle is as
small as $5\degr$. 

Runs \r{10}, \r{20}, \r{40}, and \r{m4} have peak
amplitudes $\epsilon=0.7$--$0.8$, similar to the strongest grand-design
spirals, but these large amplitudes are only present for a small fraction of
the time ($N_s\sigma_s/t_0=0.03$--0.06). The relatively small number of
high-amplitude transients in these runs is consistent with a model in
which the Galaxy has experienced several short-lived grand-design spiral
phases, possibly caused by minor mergers. In contrast, run \r{sim} has a
much larger number of weaker transients ($\epsilon=0.3$,
$N_s\sigma_s/t_0=0.17$). 

Each run was repeated seven times (realizations a-g), using different
random-number seeds to generate both the stars and the spiral transients, in
order to distinguish systematic from stochastic variations. To study
extremely long-term behavior, we also ran a realization h, in which both the
integration time and the number of transients are a factor of three larger
than in a-g.

\section{Results} \label{sec:RESULTS}

\begin{table*}
\begin{minipage}{18cm}
\caption{Simulation results and observations}
\label{r-table}
\begin{tabular}{|lcccccccc|} 
Quantity & Symbol & Run \r5 & Run \r{10} & Run \r{20} & Run \r{40} & Run
\r{m4} & Run \r{sim} & Observations \\ \hline 
Radial velocity dispersion & $\sigma_{xx}(\kms)$        & $36\pm1$  & $38\pm2$
& $40\pm 4$   & $39\pm6$  & $35\pm3$ & $40\pm3$ &36.9\\
vertex deviation & $l_v (\degr)$                        & $-3\pm4$ & $3\pm3$
& $8\pm16$  & $-8\pm 12$ & $3\pm8$ & $-6\pm11$ &10.9 \\
axis ratio of velocity ellipsoid & $\sigma_1/\sigma_2$  & $1.3\pm0.1$  &
$1.4\pm0.1$  & $1.2\pm0.1$  & $1.3\pm0.1$  & $1.4\pm0.1$ & $1.4\pm0.2$ & 1.48 \\
mean radial velocity & $\overline u (\kms)$             & $0\pm 4$  & $-1\pm3$
& $0\pm5$  & $3\pm1$  & $0\pm4$ & $0\pm8$ &$-0.5$\\
mean azimuthal velocity & $\overline v (\kms)$          & $0\pm2$ & $0\pm2$
& $-1\pm3$   & $3\pm1$ & $-2\pm3$ & $2\pm5$ &$-17.7$ \\
\hline
\end{tabular}

\medskip
Results given are the mean and standard deviation from realizations a--g, for
a population of stars with ages uniformly distributed between 0 and 10
Gyr. ``Observations'' column gives results from samples B4+GI from
\citet{deh98}, which contains stars redward of the Parenago discontinuity
($B-V>0.6$, for which the lifetime of main-sequence stars exceeds the age of
the Galaxy) and giant stars. The large negative mean azimuthal velocity in the
observations is due to the asymmetric drift, which is not present in our
simulations. 
\end{minipage}
\end{table*}

\subsection{Age-velocity dispersion relation} 
\label{sec:avr}

In \S\ref{sec:integration}, we have shown that a single set of orbit
integrations can be used to derive the \df\ for stars of all ages between 0
and $t_0$. Thus we may combine \dfs\ of different ages to produce the \df\
that corresponds to a uniform distribution of ages between 0 and 10 Gyr
(hereafter the ``combined \df''), as
would result from a uniform star-formation rate. The average moments of this
\df\ for our runs are shown in Table \ref{r-table}, along with the observed
values taken from \citet{deh98}. Of course, the close agreement of the radial
velocity dispersion $\sigma_{xx}$ in each run with the observed value of
$37\kms$ arises because the strength and number of spiral transients were
chosen to produce the correct radial dispersion.

The evolution of the principal axes ${\sigma}_1$ and ${\sigma}_2$ of the
velocity-dispersion tensor in runs \r5, \r{10}, \r{20} and \r{40} is shown in
Figure \ref{fig:age_disp}. The three rows of panels show the results for three
different realizations, i.e., three different sets of random-number seeds used
to generate both the stars and the spiral transients. In general, the results
in this and other figures are independent of the realization of the stellar
distribution, and all of the variations seen are due to changes in the
realization of the transient spiral structure---a reflection of the fact that
the Monte Carlo simulation of the stellar distribution is a numerical method
while the simulation of the distribution of spiral transients is a model of a
stochastic physical process. 

\begin{figure*}
\begin{center}
\vspace{8cm}
\includegraphics{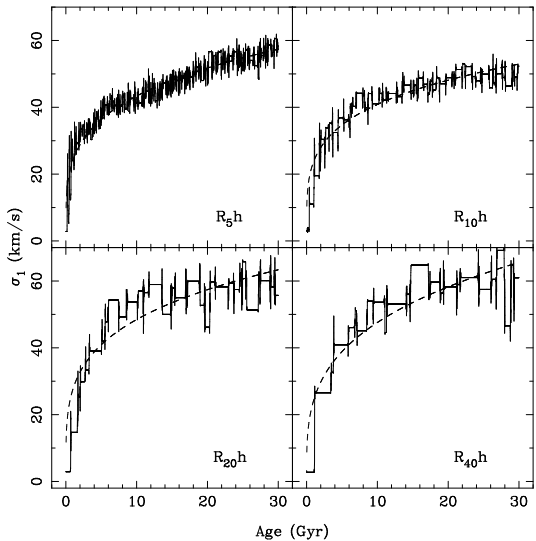}
\caption{The evolution of the \avr\ over 30 Gyr. The dashed lines show the
best-fit models of the form (\ref{eq:hist-fit}); see also Table
\ref{p-table}.}
\label{fig:age_disp_long}
\end{center}
\end{figure*}

The \avrs\ illustrate the following:

\begin{itemize}

\item Open spirals (large pitch angle) heat the disk stars more effectively
than tightly wound spirals. For example, both the number $N_s$ and the
surface-density amplitude $\epsilon$ of the spiral transients are larger in
\r5 than the other runs shown (see Table \ref{phys-param}), even though all
runs produce the same velocity dispersion by design.

\item The heating occurs in discrete steps
separated by periods of zero growth. This feature arises because we have only
a small number of spiral transients over the course of the simulation, each of
which has the same amplitude and lasts only for a characteristic time
$\sigma_s=1$ out of the total integration time of $281$. During the intervals
between the spiral transients, the \df\ is stationary so the velocity
dispersions are constant.

\item In run \r5 the growth of the dispersion slows sharply after the radial
dispersion reaches $\sim 30\kms$. This effect probably arises because the
epicycle amplitude becomes comparable to the radial spacing between the spiral
arms, so that the forces from the spirals tend to average to zero over an
epicycle oscillation. More specifically, the radial spacing of the spiral arms
is $\Delta x=2\pi R_0\tan\alpha/m$; the \rms\ epicycle amplitude
$(\overline{a^2})^{1/2}$ is given by equation (\ref{eq:epiamp}), and we expect
the spiral heating to become ineffective when the \rms\ epicycle amplitude
exceeds the peak-to-trough distance between arms, that is, when
$(\overline{a^2})^{1/2}/\Delta x\ga 0.5$. For small pitch angle $\alpha$,
\be
{(\overline{a^2})^{1/2}\over \Delta x}={m\sigma_{xx}\over 2\pi
\Omega R_0\tan\alpha}\simeq 0.50{m\over 2}{\sigma_1\over30\kms}
{5\degr\over\alpha}.
\ee

\item There are substantial differences among the realizations of a given run;
in other words there are significant stochastic effects in spiral-wave heating
(see the discussion of the exponent $p$ below for further detail). The
stochastic variations in the heating rate become smaller as the velocity
dispersion increases (see also Figure \ref{fig:age_disp_long}, which follows
the \avrs\ over an even longer time interval). 

\end{itemize}

For comparisons with theoretical models, it is useful to fit the simulated
\avrs\ with a parametrized function. A common parametrization
(e.g. \citealt{lac91}) is 
\be
\sigma_1(t)=(\sigma_0^{1/p}+Ct)^p,
\label{eq:hist-fit}
\ee
where $\sigma_0$ is the initial dispersion along the major principal axis,
fixed at $3\kms$ as in \S\ref{sec:units}.  The motivation for this form is
the assumption that the evolution of the dispersion is governed by a diffusion
equation with velocity-dependent diffusion coefficient,
\be
\frac{d \sigma_1^2(t)}{dt} = C_1 \sigma_{1}^{-q}(t), \label{eq:diff-equ}
\ee
where $C_1=2pC$ and $q=p^{-1}-2$, with $C$, $p$ and $q$ constants. For
${\sigma}_1\gg{\sigma}_0$, equation (\ref{eq:hist-fit}) implies
$\sigma_{1}(t)\propto t^p$.

The dashed lines in Figure \ref{fig:age_disp} show the results of
least-squares fits for the parameters $C$ and $p$ in equation
(\ref{eq:hist-fit}). In each case, the fitting function successfully preserves
the overall shape of the \avr\ but smooths over fluctuations due to the finite
number of transients.  We fit each of the seven realizations of each run and
obtained the values of the \avr\ exponent $p$ given in Table
\ref{p-table}.  The standard deviation in $p$
can be as large as 35\% (in run \r{20}) among realizations of a single
parameter set, and the variation due to changes in the spiral-structure
parameters is just as large. Thus, the exponent of the \avr\ is not likely to
discriminate well between heating by transient spiral structure and heating by
other mechanisms.

Figure \ref{fig:age_disp_long} shows the \avrs\ of realization h, which was
run for $t_0=30$ Gyr, three times as long as the other runs. The purpose of
this long run was to see how well the parametrization (\ref{eq:hist-fit})
worked as the heating process continued. The dashed lines show that equation
(\ref{eq:hist-fit}) still provides a reasonable fit; however, the best-fit
exponent $p$ (see Table \ref{p-table}) is systematically smaller than in the
shorter realizations a--g, except in run \r5. Therefore, equation
(\ref{eq:hist-fit}) should be considered as a fitting function valid over a
limited time interval, rather than a formula with predictive power.

\begin{figure*}
\begin{center}
\vspace{8cm}
\includegraphics{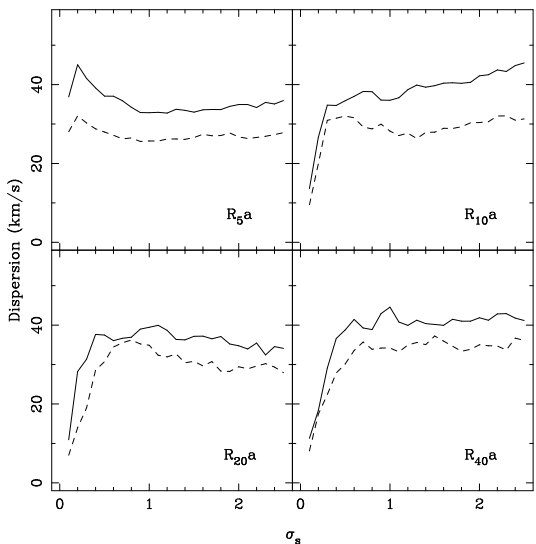}
\caption{The dependence of the velocity dispersion on the duration of the
transients, derived from the combined \df\ for a population of stars with a
uniform distribution of ages up to 10 Gyr. All of the runs plotted in a given
panel use the same defining parameters as in Table \ref{phys-param}, except
for the transient duration ${\sigma_s}$. The solid and dashed lines show
${\sigma_{xx}}$ and ${\sigma_{yy}}$, respectively. The velocity dispersion is
almost independent of the transient duration ${\sigma_s}$, except for
$\sigma_s\la0.5$.}
\label{fig:disps}
\end{center}
\end{figure*}

All of our runs have the same value of the parameter $\sigma_s$ (Table
\ref{phys-param}), which represents the duration of the spiral transients. We
have experimented with a range of values of $\sigma_s$. As shown in Figure
\ref{fig:disps}, the dispersion for the combined \df\ with a uniform
distribution of ages between 0 and 10 Gyr is almost independent of $\sigma_s$
so long as $\sigma_s\ga 0.5$. A probable explanation is that the width of the
power spectrum of the transient perturbations, given by equation
(\ref{eq:st}), is dominated by the dispersion in pattern speeds rather than
the duration of a single transient once
$\sigma_s\ga1/(2\sqrt{2}Ak_y\sigma_c)$, which occurs when $\sigma_s\ga 1.4$
for our parameters. Of course, when $\sigma_s$ becomes very large, the overlap
factor of equation (\ref{eq:overlap}), will decrease below unity and the
heating will be localized at discrete resonances.

\begin{table*}
\begin{minipage}{11cm}
\caption{Fitted exponent $p$ of \avrs\ (eq.~\ref{eq:hist-fit}).} 
\label{p-table}
\begin{tabular}{|c|c|c|c|c|c|c|c|c|c} 

    & a & b & c & d & e &f & g & a--g\footnote{The mean and standard deviation
of the values of $p$ from runs a--g.}  & h\\ \hline
\r5 & 0.31 & 0.20 & 0.22 & 0.27 & 0.27 & 0.22 & 0.31 & 0.25$\pm$0.05 & 0.24\\
\r{10} &  0.30 &  0.34 & 0.37 & 0.32 & 0.24 & 0.24  & 0.25 & 0.29$\pm$0.05 & 0.25\\
\r{20} &  0.33 & 0.46 & 0.35 & 0.59 & 0.29 & 0.76 & 0.57 & 0.48$\pm$0.17 & 0.29\\
\r{40} &  0.37 & 0.50 & 0.38 & 0.53 & 0.36 & 0.51 & 0.40 & 0.44$\pm$0.08 & 0.23\\
\r{m4}  & 0.48 & 0.28 & 0.26 & 0.44 & 0.24 & 0.53 & 0.26 & 0.36$\pm$0.12 &  -\\
\r{sim} & 0.40 & 0.43 & 0.55 & 0.40 & 0.39 & 0.41 & 0.66 & 0.46$\pm$0.10 &  - \\
\end{tabular}
\end{minipage}
\end{table*}

\subsection{The velocity ellipsoid}

\begin{figure*}
\begin{center}
\vspace{15cm}
\includegraphics{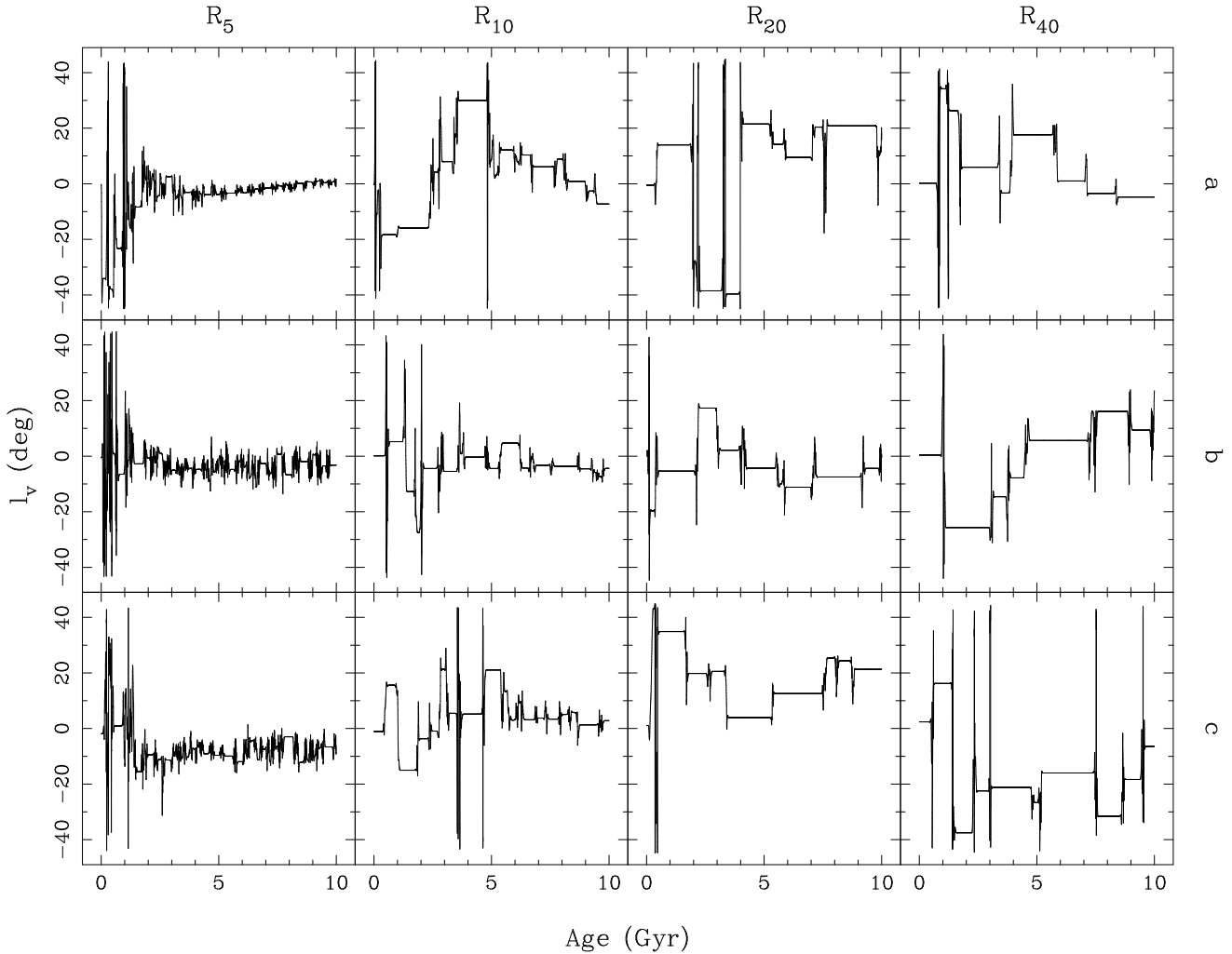}
\caption{The evolution of the vertex deviation $l_v$ (eq. \ref{eq:vert}). The
three rows represent different realizations.} 
\label{fig:age_vertex}
\end{center}
\end{figure*}

The time evolution of the vertex deviation is shown in Figure
\ref{fig:age_vertex}. Not surprisingly, the stochastic variations in the
vertex deviation are larger for runs \r{10}, \r{20} and \r{40}, which have
larger pitch angle and fewer transients, and the fluctuations in the vertex
deviation decline with time as the growth rate of the velocity dispersion
declines. The small but non-zero vertex deviations seen near the end of the
runs are roughly consistent with the vertex deviation $\sim 10\degr$ seen in
the old stellar population in the solar neighbourhood. There is no evidence of
a systematic preference for positive or negative vertex deviations for old
stars.

\begin{figure*}
\begin{center}
\vspace{15cm}
\includegraphics{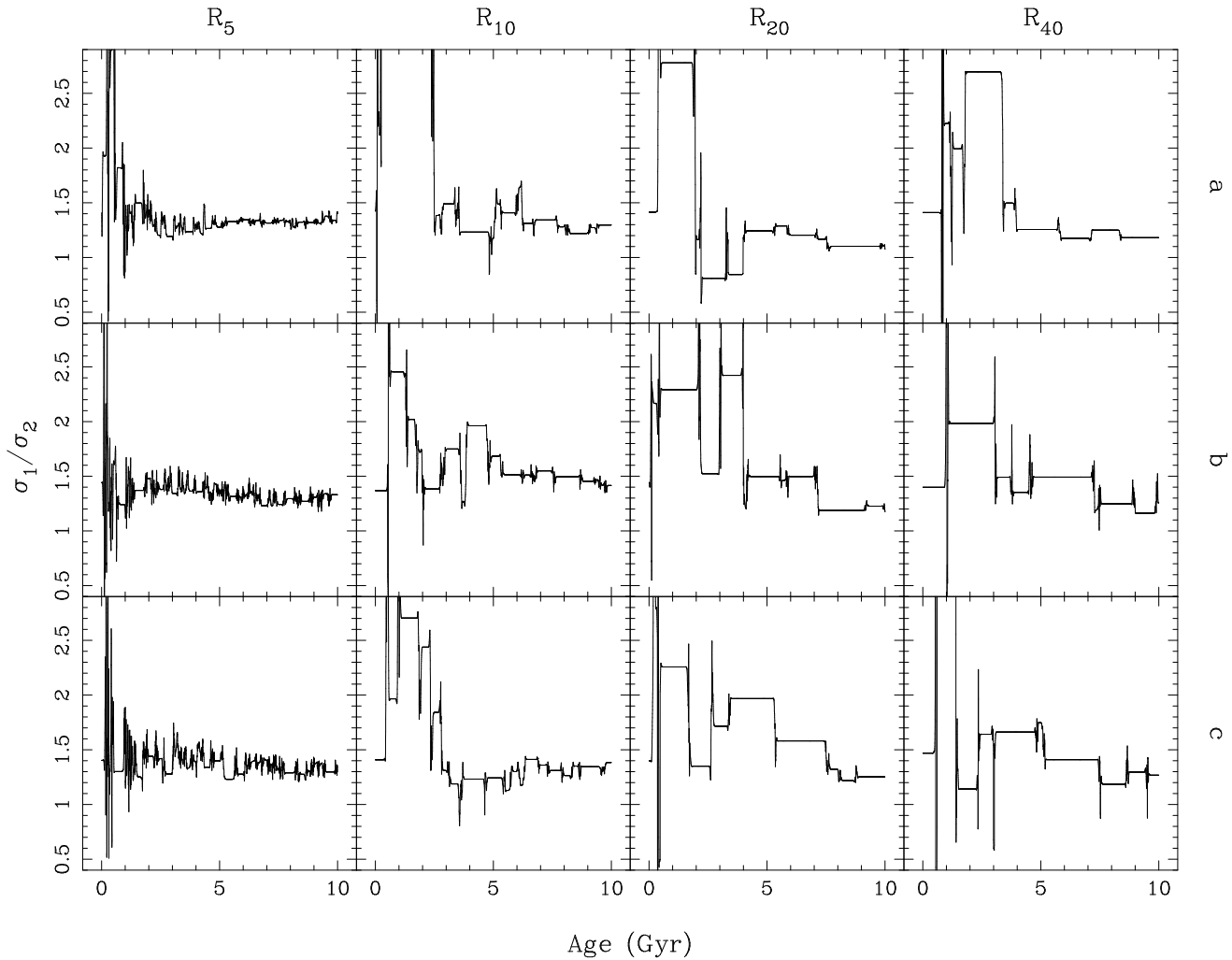}
\caption{The evolution of the ratio $\sigma_1/\sigma_2$ between the major and
minor principal axes of the velocity-dispersion tensor. In a steady-state
axisymmetric galaxy with a flat rotation curve, the axis ratio should be
$2^{1/2}=1.414$.} 
\label{fig:age_ratio}
\end{center}
\end{figure*}

Figure \ref{fig:age_ratio} shows the evolution of the ratio of the major and
minor principal axes of the velocity ellipsoid. As in the case of the vertex
deviation, the fluctuations are much larger in the runs with larger pitch
angle. In a steady-state axisymmetric galaxy with a flat rotation curve, the
axis ratio should be
$\sigma_1/\sigma_2=[\Omega/(\Omega-A)]^{1/2}=2^{1/2}=1.414$
(eq. \ref{eq:axis-ratio}), and the observed axis ratio gradually approaches
this value as the runs progress.

\begin{figure*}
\begin{center}
\vspace{15cm}
\includegraphics{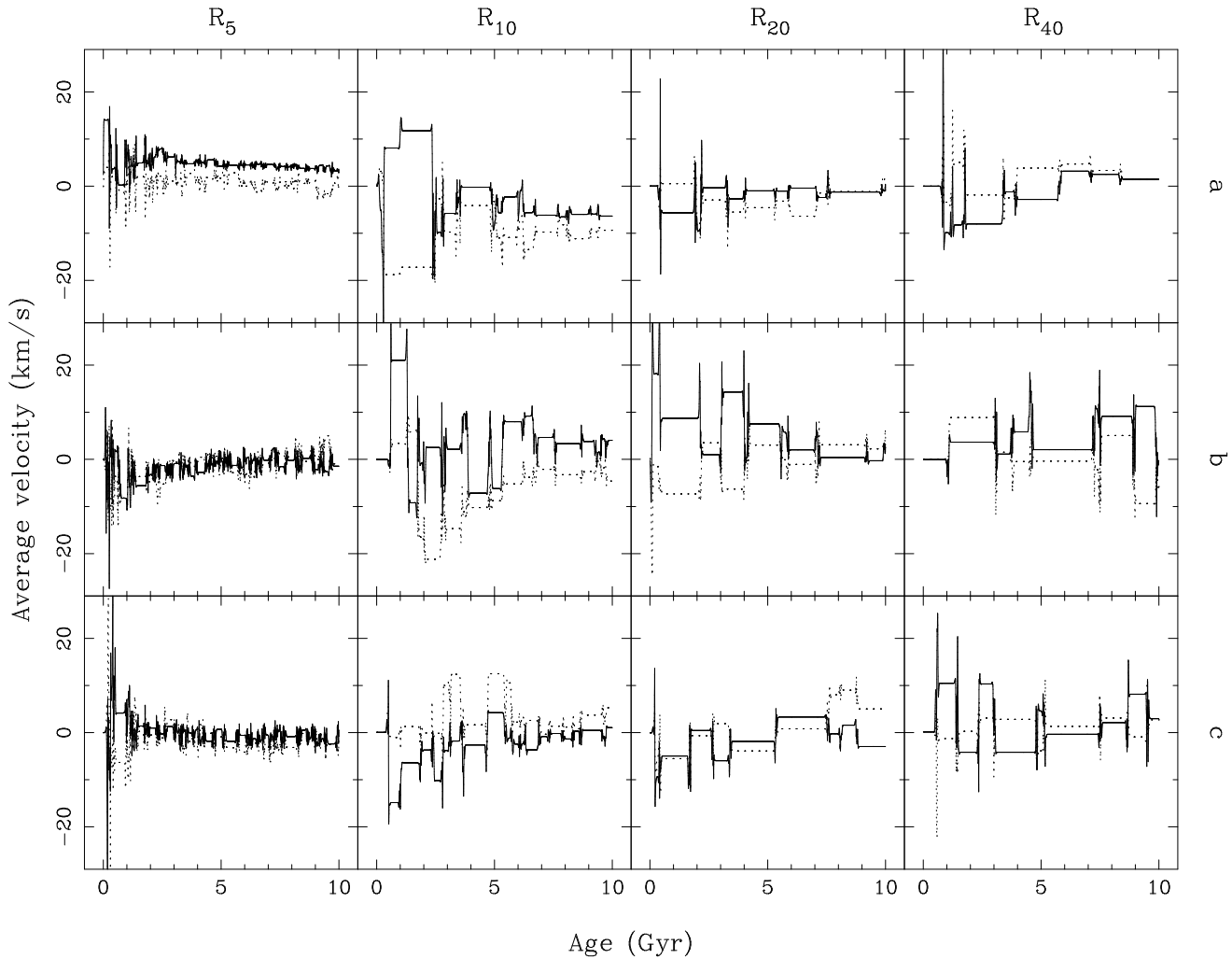}
\caption{The evolution of the mean velocities $\overline u=-\overline 
v_x$ (solid line) and $\overline v=\overline v_y$ (dotted line).}
\label{fig:age_meanuv}
\end{center}
\end{figure*}

Figure \ref{fig:age_meanuv} shows the evolution of the mean radial and
azimuthal velocity in the solar neighbourhood. There are significant
fluctuations, up to $10\kms$ for the larger pitch angles, even in the oldest
stellar populations. These fluctuations limit the accuracy with which we can
construct axisymmetric models of local Galactic kinematics. 

We have also plotted the axis ratio of the velocity ellipsoid against the
vertex deviation but have found no systematic correlation between these two
quantities. 

The parameters of the \df\ for a population of stars with a uniform
distribution of ages between 0 and 10 Gyr are shown in Table
\ref{r-table}. They are all consistent with the observations, except for the
mean azimuthal velocity $\overline v$. The discrepancy in $\overline v$
reflects the fact that our simulations do not show asymmetric drift because of
the approximations inherent in the sheared sheet [eqs.\ (\ref{eq:y-frame})
with the perturbing potential $\Phi=0$ are symmetric under reflection through
the origin].

\begin{figure*}
\begin{center}
\vspace{15cm}
\includegraphics{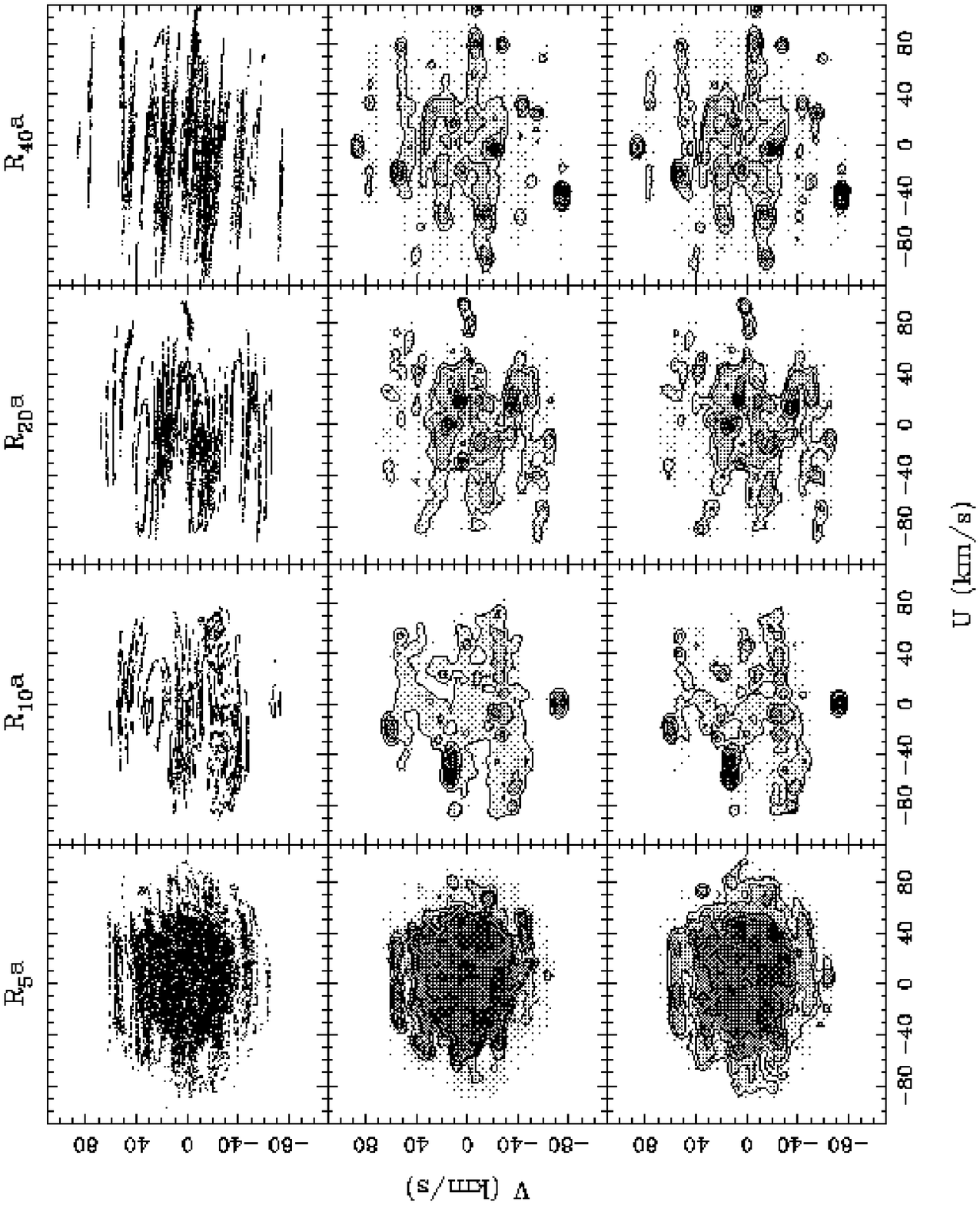}
\caption{The velocity \df\ for stars of age 5 Gyr, from runs \r{5}a, \r{10}a,
\r{20}a, 
and \r{40}a. The top row shows locations from our Monte Carlo sampling of
velocity space that yield nearly circular orbits (epicycle energy
$<3\sigma_0^2$, $\sigma_0=3\kms$) at the formation epoch. The second and third
rows show contour plots of the \df\ smoothed with observational errors
$\sigma_{\rm ob}=3\kms$, from two realizations with the same spiral-structure
parameters but different sets of stars. The differences between the second and
third rows are a measure of the uncertainties in the \dfs\ arising from our
numerics. The contours are at $0.2,0.4,\ldots,1.0$ times the maximum value of
the \df\ in each panel.}
\label{fig:contour}
\end{center}
\end{figure*}

\subsection{Shape of the velocity distribution\label{sec:df}}

Figure \ref{fig:contour}\ shows the \df\ for stars with age 5 Gyr, from runs
\r{5}a, \r{10}a, \r{20}a and \r{40}a.  We present the \df\ in two forms: (i)
we display all of the points from our sampling of velocity space at the
present epoch that yield nearly circular orbits at the formation epoch
(i.e. those with epicycle energy $\le3\sigma_0^2$, $\sigma_0=3\kms$). (ii) We
show contour plots of the smooth \df\ that is expected with Hipparcos
observational errors at $\sim 100\pc$ (see eq.\ \ref{eq:error}), for two
different realizations of the stellar distribution and the same realization of
the spiral transients. The difference between these two realizations (the
second and third rows) is an indication of the numerical noise in
our simulations, which is small.

The simulated \dfs\ in the second and third row of panels do not resemble the
Gaussian or Schwarzschild \df\ (eq.\ \ref{eq:sss}). They are much lumpier, and
the lumps are distributed more-or-less uniformly over a large area in velocity
space, rather than being concentrated at zero velocity. Qualitatively, the
Schwarzschild \dfs\ resemble a contour map of a single large mountain, whereas
the simulated \dfs\ resemble an entire mountain range. The ``lumpiness'' or
degree of substructure in the \df\ is a strong function of the pitch angle of
the spiral arms, and in the range of pitch angles $10\degr$--$20\degr$ that
observations suggest for our Galaxy the \dfs\ are highly irregular. 

The \dfs\ in Figure \ref{fig:contour} exhibit prominent streaks in the
direction $v=v_y=$constant, particularly at the larger pitch angles. The
explanation of these streaks is straightforward, and dates back to
\citet{woo61}. It is straightforward to show from equations (\ref{y-epi}) that
for stars in the solar neighbourhood ($x=y=0$) the radial and tangential
velocities $(u,v)$ are related to the guiding-center coordinates
$(x_g,y_g)$ by 
\be
(x_g,y_g)={1\over 2(\Omega-A)}(v,u).
\label{eq:toomre}
\ee
Suppose that scattering from a spiral transient leads to an enhancement in the
density of stars over the range of guiding-centre radii $x_g\to x_g+\Delta
x_g$. Since the guiding-centre velocity is $\dot y_g=-2Ax_g$
(eq. \ref{y-epi}), the stars will slowly spread out into an extended streamer
in physical space; after time $t$ the streamer will have width of order the
epicycle amplitude and length $\Delta y\simeq 2A\Delta x_gt$. The stars found
at a given location along the streamer (for example the solar neighbourhood)
will thus have very similar values of $x_g$ and thus of $v$. In other words, a
tidal streamer will present itself in the local velocity distribution as a
streak with constant $v$.

The 1:1 relation between $(u,v)$ and $(y_g,x_g)$ implied by equation 
(\ref{eq:toomre}) implies that the structure seen in Figure \ref{fig:contour} 
is closely related to the guiding center correlations discussed by
\citet{tk91}. 

\citet{shc99} find that the observed \df\ in the solar neighbourhood
exhibits parallel ``branches'' of enhanced density in velocity space. It is
tempting to identify these with the streaks predicted by \citet{woo61} and
seen in Figure \ref{fig:contour}. However, the branches found by \citet{shc99}
are tilted with respect to the expected orientation $v=$constant; they find
instead that $v=\hbox{constant}-0.47u$. Thus the correct interpretation of
these branches is unclear: they may arise from some physical effect we have
not modeled, or they may be artifacts arising from our predisposition to find
order in random patterns. Note, for example, that parallel tilted ``branches''
appear to be present in the contour plots of the \df\ in run
\r{20}a shown in Figure \ref{fig:contour}, with a slope about half of that
claimed by Skuljan et al. 

The principal conclusion from Figure \ref{fig:contour} is that the lumps in
the simulated \dfs, at least in the simulations with pitch angles
$\ga10\degr$, bear a striking resemblance to the moving groups in the solar
neighbourhood (see for example \citealt{deh98}, \citealt{shc99}, or Chereul et
al.\ 1998, 1999), which strongly suggests that at least the older moving
groups are a consequence of the same transient spiral structure that heats the
stellar distribution. Thus we suggest that old moving groups arise from
fluctuations in the gravitational potential rather than fluctuations in the
star-formation rate, as has usually been assumed.  This result was anticipated
to some extent by \citet{kal91} and Dehnen (1999, 2000), who argued that some
of the most prominent substructure in the solar neighbourhood \df\ arose from
the outer Lindblad resonance with the Galactic bar. The distinction is that
they focused on a single steady bar-like potential whereas we have studied
stochastic spiral transients.

\subsection{Evolution of moving groups}
\label{sec:obsdf}

\begin{figure*}
\begin{center}
\vspace{13cm}
\includegraphics{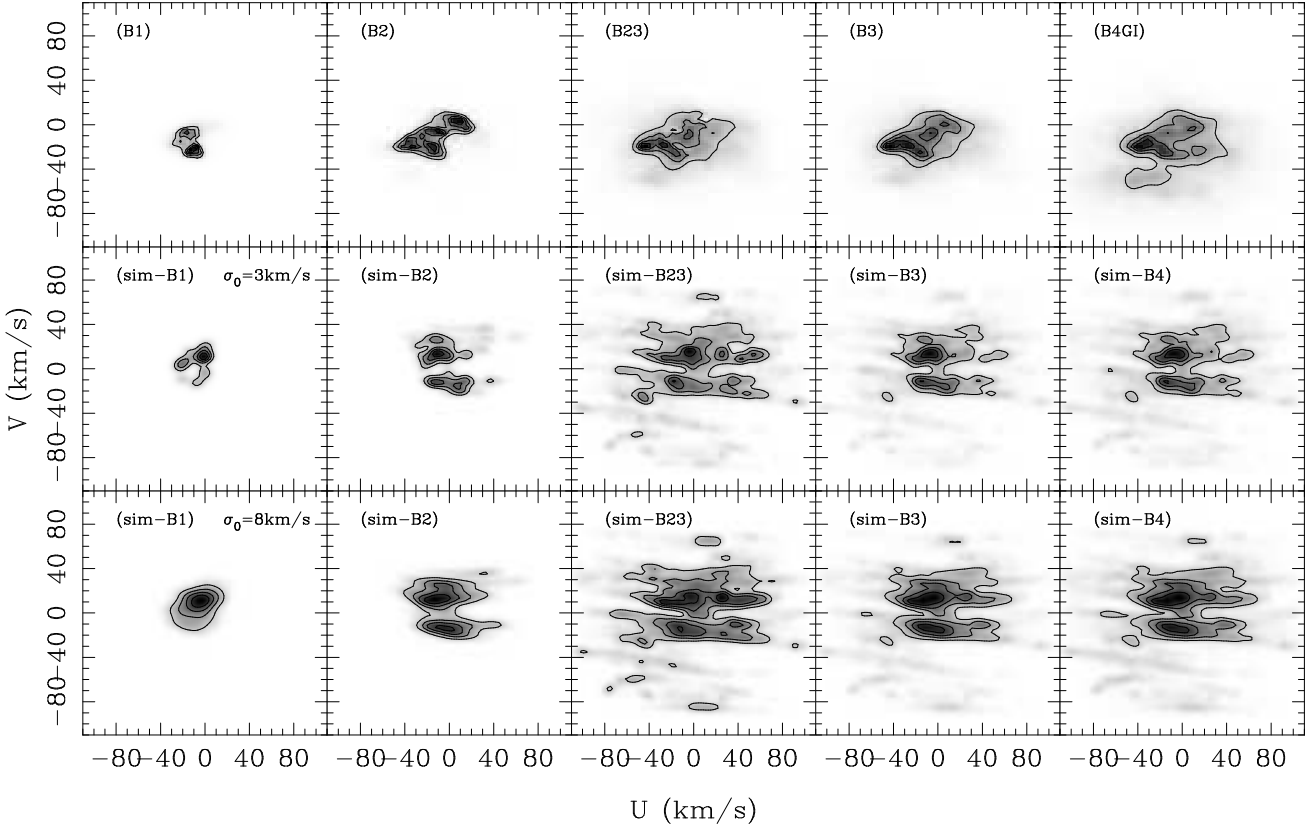}
\caption{The observed \df\ as a function of stellar age \citep{deh98},
compared to simulated \dfs\ from run \r{sim}a. From left to right, the samples
represent stars younger than 0.4 Gyr, stars younger than 2 Gyr, stars from
2--8 Gyr in age, stars younger than 8 Gyr, and stars of all ages (see text for
further details).  The
bottom row is from a run with the same parameters as \r{sim}a except that
initial dispersion ${\sigma_0}$ is $8\kms$ rather than $3\kms$.}
\label{fig:simdf}
\end{center}
\end{figure*}

One of the striking features of the solar neighbourhood \df\ is that the same
moving groups appear to be present in stellar populations of different ages
(see Table 2 of \citealt{deh98} or the top row of Figure
\ref{fig:simdf}). This result is difficult to understand in the traditional
view that moving groups arise from inhomogeneous star formation. Thus it is
worthwhile to explore the temporal evolution of the \df\ and the survival of
moving groups in our simulations.

Our standard four simulations, \r5, \r{10}, \r{20}, and \r{40}, are not ideal
for this task. As we have seen, run \r5\ has relatively weak moving
groups and unrealistic parameters. The other runs show a strong moving group 
at the origin in the combined \df\, which is not
present in the observed \df. This moving group arises because the typical
interval between spiral transients in these simulations is 0.5--1 Gyr (Table
\ref{phys-param}). Thus in most of these runs there has been no spiral
transient in the last $\sim 0.5$ Gyr, and all of the stars formed since the
last transient still have their birth epicycle energy, which is nearly zero.
Moreover, these runs feature a relatively small number of strong spirals,
which hit the disk very infrequently but hard. They may therefore overestimate
the persistence of moving groups.

To address these concerns, we have run another simulation \r{sim}. In this
run, we increase the number of spirals to $N_s=48$ and set $\epsilon=0.31$.
We have also adjusted the central times of the spiral transients as follows:
(1) We have set the central time $t_s$ of the most recent transient to be the
present time $t_0$; (2) Because Dehnen's subsample B1, in which the oldest
stars have age 0.4 Gyr, already shows two prominent moving groups (top left
panel of Figure \ref{fig:simdf}), we have set the central time of another
transient to $t_0-0.2$ Gyr; (3) The other central times are chosen randomly
between $t_s=0$ and $t_s=t_0-0.2$ Gyr. This strategy suppresses the strong
moving group at the origin and makes the other features in the \df\ more
prominent. 

\begin{figure*}
\begin{center}
\vspace{15cm}
\includegraphics{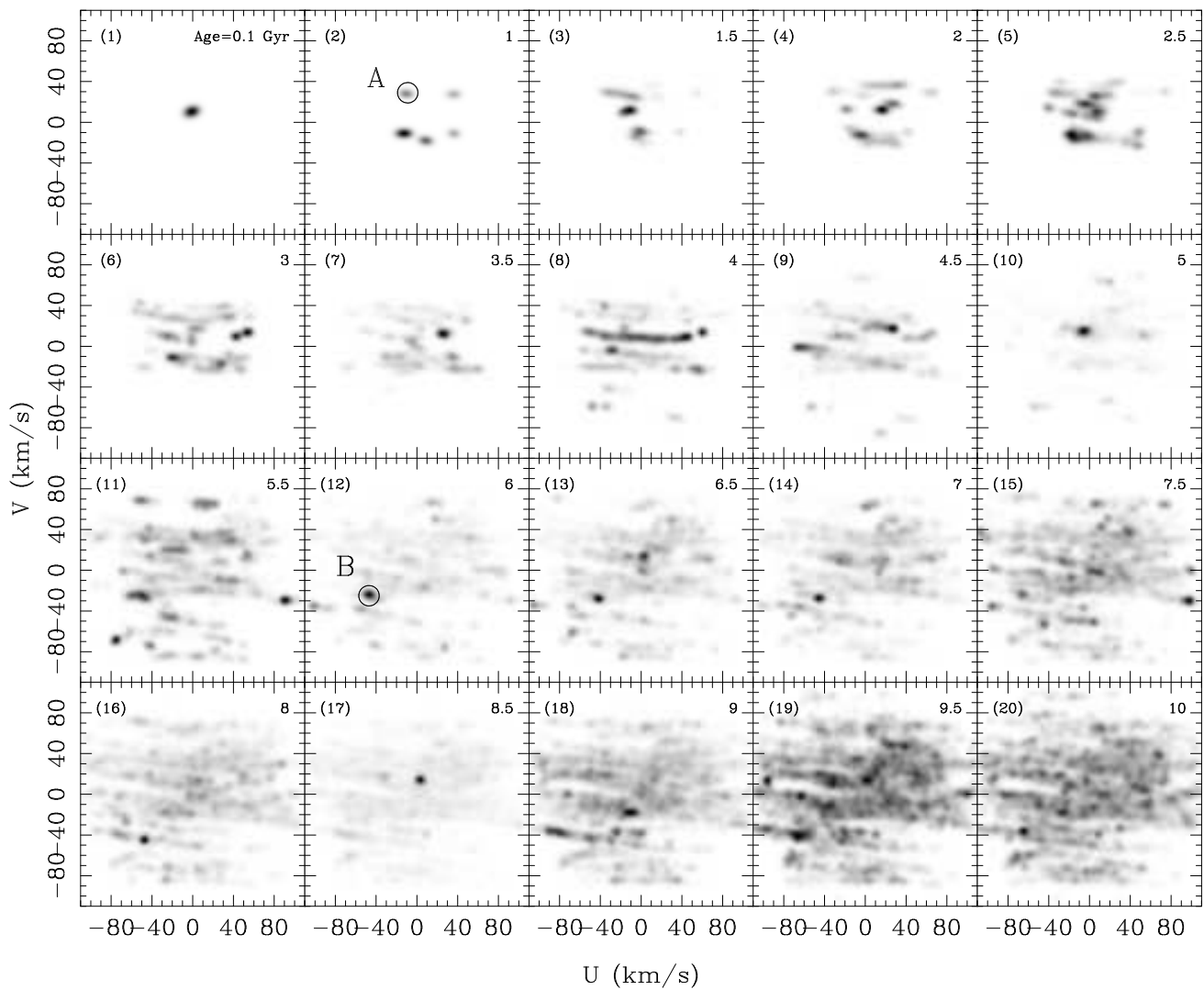}
\caption{The evolution of the velocity distribution as a function of age, from
run \r{sim}a. Each panel is labeled with the age in Gyr. The gray scale in
each panel is relative to the maximum density in that panel. Notice that the
moving group B can be seen at least from panels 7 to 13, corresponding to ages
of 3.5--6.5 Gyr.}
\label{fig:dfevolution}
\end{center}
\end{figure*}

Figure \ref{fig:dfevolution} shows the velocity distribution in run \r{sim}a
as a function of stellar age. ``Moving groups'' (i.e. the same local
concentration of stars in velocity space) can be seen among stars of any age,
even the oldest stars.  Generally with increasing age there are more but
weaker moving groups, however, a strong ``moving group'' can be present even
in a very old sample (see panel 17, 8.5 Gyr).  Notice that the same ``moving
group'' can be present in stars with a wide range of ages; for example, the
moving group ``B'' in panel 13 can be traced from panels 11 to 14, a range of
at least 1.5 Gyr in age. This result is consistent with the observation that
the same moving group is seen in observational sub-samples of different age
(e.g., \citealt{deh98}, Table 2).

In Figure \ref{fig:simdf} we show the observed and simulated \dfs, divided
into age bins. The observed \df, shown in the top row, is taken from
\citet{deh98}. The panels represent subsamples determined by color cuts, which
should correspond to age cuts. The maximum ages in the three subsamples B1,
B2, and B3 (the first, second, and fourth panels) are 0.4 Gyr, 2 Gyr, and 8
Gyr (Table 1 of \citealt{deh98}). The sample B4GI (fifth panel) is a
combination of red ($B-V>0.6$) main-sequence stars and giant stars, which
should have a uniform distribution of ages if the star-formation rate has been
uniform. The third panel, labeled B23, is obtained by subtracting the scaled
\df\ of B2 from the \df\ of B3, in order to represent approximately the \df\
of stars with ages 2--8 Gyr. 

In the bottom two rows of Figure \ref{fig:simdf} we show the simulated \dfs\
from two runs: run \r{sim}, and a modified version of \r{sim} in which the
width $\sigma_0$ of the epicycle energy at formation (eq. \ref{eq:Schwarz}) is
increased from $3\kms$ to $8\kms$. The different panels show the \dfs\
corresponding to the same age ranges as in the subsamples of the data. The
simulated \dfs\ reproduce most of the prominent features of the observed
\dfs: (1) The stars clump together in moving groups. 
(2) The number of moving groups increases as the mean stellar age increases,
from left to right in the figure. (3) The same moving groups are present in
stars with a wide range of ages. (4) The moving groups appear to be elongated
along lines of constant $v$, although this effect is significantly stronger in
the simulations than the observations.

Increasing the initial epicycle energy (compare the second and third row of
Figure \ref{fig:simdf}) spreads out the moving groups somewhat,
and enhances their elongation along lines of constant $v$, but they remain
prominent; the moments of the \df\ given in Table \ref{r-table} are almost
unchanged.

\begin{figure*}
\begin{center}
\vspace{9cm}
\includegraphics{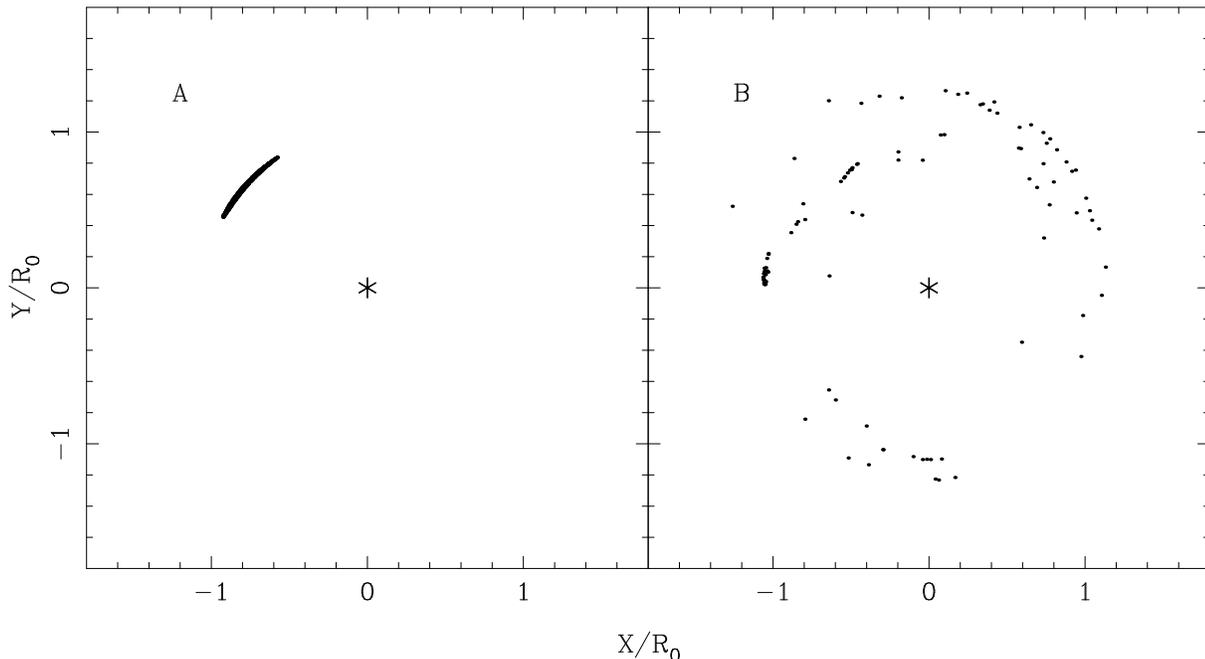}
\caption{The birthplaces of stars in moving groups ``A'' and ``B'' in Figure
\ref{fig:dfevolution}.  The star shows the Galactic centre. The stars in a
moving group are not born at a single location. The locations are obtained
from the coordinates of the stars in the sheared sheet by equating the polar
coordinates $R/R_0$ and $\phi$ to $1+x$ and $y$, respectively.}
\label{fig:f_place}
\end{center}
\end{figure*}

In Figure \ref{fig:f_place} we plot the birthplaces of groups ``A'' and ``B''
shown in panels 2 and 12 of Figure \ref{fig:dfevolution}.  The stars in the
younger group, A (age 1 Gyr), come from a relatively small arc, while the
stars in the older moving group B (age 6 Gyr) were born with a wide range of
Galactic azimuths. In contrast, Eggen's model for moving groups predicts that
the stars in a group should be born in a small region and thus should have the
same age and metallicity.

We also ran a modified version of \r{20}, which has the same parameters as
\r{20} but adjusts the central times of the transients as described above for
\r{sim}. This run gives more prominent moving groups than \r{sim}, and the
groups last for a longer time (up to 3.5 Gyr). This result suggests that a
small number of strong spiral transients tend to produce strong moving groups
and a lumpy \df, while a large number of weak transients tend to produce weak
or undetectable moving groups and a smooth \df. Therefore, the observed moving
groups suggest that at most a few tens of spiral transients have affected the
stars in the solar neighbourhood.

We also looked for moving groups in run \r{m4}, which has the same parameters
(including the random number seeds) as the modified run \r{20}, except that
the spirals have four arms\footnote{It is striking that run \r{m4} yields
almost the same velocity dispersion as run \r{20} (Table \ref{r-table}). The
reason is as follows: consider a simulation with azimuthal wavenumber $m=2$
that yields a power-law age-velocity dispersion relation of the form
(\ref{eq:hist-fit}), $\sigma(t)=(Ct)^p$. We expect that the constant $C$ is
proportional to the diffusion coefficient, and that this in turn is
proportional to the mean-square fluctuating force $F^2\equiv (k\Phi_s)^2$;
thus $C=cF^2$ where $c$ is a constant and the age-velocity dispersion relation
can be written as $\sigma(t)=F^{2p}(ct)^p$. According to the scaling relations
at the end of \S\ref{sec:spiral}, a simulation with azimuthal wavenumber $m$
is equivalent to one with $m'=f m$ (here $f=2$) if $\sigma'=\sigma/f$ and
$F'=F/f$. Thus $\sigma'(t)=F^{2p}(ct)^p/f=f^{2p-1}(F')^{2p}(ct)^p$. Since the
exponent $p\simeq 0.5$ for run \r{20}, $f^{2p-1}\simeq 1$ and we expect the
age-velocity dispersion relations for runs \r{20} and \r{m4} to be nearly the
same.}.  We found that substructure in the \df\ in run \r{m4} was much weaker
than in the modified run \r{20}, presumably because the wavelength $2\pi/k$ of
the arms is smaller so the \df\ is more thoroughly mixed for a given amount of
heating.  The absence of substructure suggests that transient spirals in the
Galaxy may be predominantly two-armed rather than four-armed. Further
exploration of the effects of the spiral-structure parameters on the structure
of the local \df\ is obviously worthwhile.

We have also looked for substructure in the velocity \df\ in runs in which the
duration $\sigma_s$ of the spiral transients was varied (Figure
\ref{fig:disps}). We found that the strength of the substructure declines as
$\sigma_s$ increases; in other words, spiral transients with duration much
longer than $\Omega^{-1}$ tend to heat the disk without generating moving
groups. 

We have also investigated heating by \gmcs\ rather than spiral structure, and
find that this process does not yield significant substructure in the \df\
\citep{des00}. Thus the observed strong substructure suggests that spiral
structure rather than \gmcs\ is the dominant heating mechanism in the solar
neighbourhood, a conclusion reached by others from independent arguments (see
\S\ref{sec:intro}).  

\section{Inhomogeneous star formation}\label{sec:inhomo}

So far in this paper we have assumed that the star-formation rate is uniform
in both space and time, and that both the \avr\ and the structure of the
velocity distribution in the solar neighbourhood are due to scattering by the
large-scale potential fluctuations associated with transient spiral arms. The
traditional view, which we explore briefly in this Section, is that the
structure in the local velocity distribution results from spatial and temporal
variations in the star-formation rate. Of course, some elements of both
pictures must be correct, since we know both that spiral arms are present and
that star formation is concentrated in \gmcs.

The observed \avr\ requires heating by some mechanism---whether spiral arms,
\gmcs, or one of the other mechanisms discussed in \S\ref{sec:intro}. However,
in this Section we ignore the details of this heating, to focus on the
properties of the irregularities in the velocity distribution that are due to
inhomogeneous star formation alone. We consider the following crude model:

\begin{enumerate}

\item  All stars form in clusters of $N$ stars, of limited spatial extent.

\item The cluster dissociates after the stars form. When the cluster
dissociates, the \df\ of its stars is Gaussian in both position and velocity,
with dispersions $\sigma_q$ and $\sigma_p$ respectively. The stars do not
interact with one another once the cluster has dissolved; thus, their motion
is governed by equations (\ref{y-epi}). Typical values of $\sigma_q$ and
$\sigma_p$ are a few pc or $\kms$.

\item  The heating mechanism does not disrupt or spread the
phase-space stream of stars from the dissolved cluster (any such effects will
make many of the conclusions below even stronger). We ignore the details of
the heating mechanism, and account for heating by simply assuming that the
dissociated cluster is born at position $(x_0,y_0)$, on a non-circular orbit
with velocities $u_0$ and $v_0$. Thus, the initial \df\ of the dissociated
cluster is
\be
f_0(\bfz)={N\over(2\pi\sigma_p\sigma_q)^2}\exp\left[-\half(\bfz-\bfz_0)
\bfQ(\bfz^T-\bfz_0^T)\right].
\label{eq:finit}
\ee
where  $\bfz=(x,y,u,v)$, $T$ denotes the transpose, and
\be
\bfQ=\left(\begin{array}{cccc}
\sigma_q^{-2} & 0 & 0 & 0 \\
0 &\sigma_q^{-2} & 0 & 0  \\
0 & 0 & \sigma_p^{-2} & 0 \\
0 & 0 & 0 & \sigma_p^{-2} \\
\end{array} \right).
\ee
The covariance matrix is
\be
\langle (\bfz^T-\bfz_0^T)\otimes(\bfz-\bfz_0)\rangle=\bfQ^{-1},
\label{eq:pdefi}
\ee
where the average $\langle\cdot\rangle$ is over the \df\ $f_0(\bfz)$. 
\end{enumerate}  

The trajectories (\ref{y-epi}) can be rewritten in the form \citep{asi99}
\be
\bfz(t)=\bfA(t)\bfz(0),
\label{eq:evolve}
\ee
where $t$ is the time elapsed since the dissociation of the cluster, and
(recall that $u=-\dot x$)
\be
\bfA(t)=\left(\begin{array}{cccc}
{\Omega-A\cos\tau\over\Omega-A} & 0 & -{\sin\tau\over\kappa} &
{1-\cos\tau\over 2(\Omega-A)} \\
{2A\Omega(\sin\tau-\tau)\over\kappa(\Omega-A)} & 1 &
{2\Omega(1-\cos\tau)\over\kappa^2} & 
{\Omega\sin\tau-A\tau\over\kappa(\Omega-A)} \\
-{A\kappa\sin\tau\over \Omega-A} & 0 & \cos\tau & -{\kappa\sin\tau\over
2(\Omega-A)} \\
{2A\Omega(\cos\tau-1)\over\Omega-A} & 0 &
{2\Omega\sin\tau\over\kappa} & {\Omega\cos\tau-A\over \Omega-A}
\\ \end{array} \right).
\label{eq:eqmatrix}
\ee
Here $\tau\equiv\kappa t$. The matrix $\bfA(t)$ has the following properties:
$\bfA(0)=\bfI$ (by definition), $|\bfA(t)|=1$ (by Liouville's theorem), and
$\bfA^{-1}(t)=\bfA(-t)$ (by time-reversibility).

Since phase-space density is conserved along the particle trajectories, the
initial \df\ is Gaussian (eq.~\ref{eq:finit}), and $\bfz(t)$ is a linear
function of $\bfz(0)$, the \df\ of the dissociated cluster will remain 
Gaussian for all time. The centroid of the Gaussian will be
$\bfz_c(t)=\bfz_0\bfA^T(t)$ and the covariance matrix will be
\be
\bfC(t)\equiv \langle [\bfz^T-\bfz_c^T(t)]\otimes[\bfz-\bfz_c(t)]\rangle=
\bfA(t)\bfQ^{-1}\bfA^T(t).
\label{eq:pdef}
\ee

The disrupted cluster is greatly elongated in the azimuthal
direction. Therefore to estimate the maximum number of stars from a cluster
that could be visible in the solar neighbourhood, it is useful to compute the
linear number density at time $t$, $n(y,t)=\int f(\bfz,t)dx\,d\bfv$, where
$\bfz=(x,y,u,v)=(\bfx,\bfv)$. Since the
\df\ is Gaussian, the linear density must also be Gaussian,
\be
n(y,t)={N\over (2\pi\sigma_y^2)^{1/2}}\exp\left[-(y-y_c)^2/2\sigma_y^2
\right],
\ee
where 
\be
\sigma_y^2=\langle[y-y_c(t)]^2\rangle=C_{22}(t);
\ee
at large times, $\kappa t\gg1$, we have \citep{asi99}
\be
\sigma_y\simeq {4A\Omega
t\over\kappa^2}(\sigma_p^2+4\Omega^2\sigma_q^2)^{1/2}.
\ee
For a typical star-forming cluster, $\Omega\sigma_q\ll\sigma_p$
($\Omega\sigma_q=0.03\kms(\sigma_q/1\pc)$ compared to typical values
$\sigma_p=1\kms$). Thus we may drop the term involving $\sigma_q$, so 
the maximum linear density from a disrupted cluster of age $t$ is 
\be
n_{\rm max}(t)={N\over (2\pi)^{1/2}\sigma_pt}\left(\kappa^2\over
4A\Omega\right), 
\ee
where the factor in brackets is unity for a flat rotation curve. In the solar
neighbourhood, 
\be
n_{\rm max}(t)=4\times10^{-5}\pc^{-1}
N\left(1\kms\over\sigma_p\right)\left(10\Gyr\over t\right).
\ee
With the same assumptions, the \rms\ spatial extent of the cluster in the 
radial direction is 
\be
\sigma_x={\sigma_p\over\Omega}\left[\half\sin^2\kappa t+(1-\cos\kappa
t)^2\right]^{1/2};
\ee
averaging over the epicycle period we have $\sigma_x=({7\over
4})^{1/2}\sigma_p/\Omega=50\pc(\sigma_p/1\kms)$.

Now suppose that our survey includes all stars in the solar neighbourhood to a
distance $L$. Then if $L\ga \sigma_x$ we can detect up to $2n_{\rm max}L$
stars in the moving group, while if $L\ll\sigma_x$ we will detect up to 
$(\pi/2)^{1/2}n_{\rm max}L^2/\sigma_x$ stars in
the group (these upper limits are achieved if the Sun lies exactly on the
centroid $x_c(t)$). Extrapolating these formulas until they meet, we find that
the upper limit to the number of detected stars in the moving group is given
by
\be
n_{\rm mg}<8\times10^{-3}N{L\over100\pc}
{1\kms\over\sigma_p}{10\Gyr\over t}
\ee
for $L>80\pc(\sigma_p/1\kms)$, and
\be
n_{\rm mg}<1.0\times10^{-2}N\left(L\over100\pc\right)^2
\left(1\kms\over\sigma_p\right)^2{10\Gyr\over t}
\ee
for $L<80\pc(\sigma_p/1\kms)$. The sample of \citet{deh98} was limited to
distances $L\simeq 100\pc$ (the median parallax error in the Hipparcos catalog
is $\sigma_\pi=1$ mas and Dehnen's sample is restricted to stars with
$\sigma_\pi/\pi\le0.1$). If we assume that at least $n_{\rm mg}=50$ stars are
needed to define a moving group, then any such group identified in Dehnen's
catalog must have come from a cluster with initial population
\be
N\ga 6\times 10^3{\sigma_p\over 1\kms}{t\over
10\Gyr}\quad\hbox{if $\sigma_p<1.3\kms$},
\ee
or
\be
N\ga 5\times 10^3\left(\sigma_p\over 1\kms\right)^2{t\over
10\Gyr}\quad\hbox{if $\sigma_p>1.3\kms$}.
\ee
These are conservative lower limits, since (i) we have neglected the spreading
of the stream due to interactions with small-scale fluctuations in the
gravitational potential, such as \gmcs; (ii) we have assumed that the radial
coordinate of the stream of stars is centred precisely on the Sun. 

Infrared observations show that most star formation occurs in embedded star
clusters within \gmcs. These clusters have masses in the range
$50$--$10^3M_\odot$ \citep{ll03} and virial dispersions $\sigma_p\sim 1\kms$
and thus are too small to contribute detectable moving groups with an age
exceeding $\sim 1\Gyr$. An alternative is to suppose that the dissociated
cluster is the \gmc\ itself. In this case $\sigma_p\sim 5\kms$ which requires
$N\ga 1.3\times 10^5$; given that the largest \gmcs\ have gas masses $\sim
10^5$--$10^6M_\odot$ we would require extremely high star-formation efficiency
to produce a sufficient number of stars. In other words it is difficult to
find a plausible astrophysical source for the star-forming regions which are
supposed to be the progenitors of the old moving groups.

\section{Migration} \label{sec:ORBIT}

\citet{sb02} discuss the dynamics of radial migration due to transient spiral
arms, and the relationship between radial migration and disk heating. They
estimate that old stars formed in the solar neighbourhood should now be
scattered nearly uniformly throughout the annulus from $R_0\pm4\kpc$. 

\begin{figure*}
\vspace{11cm}
\includegraphics{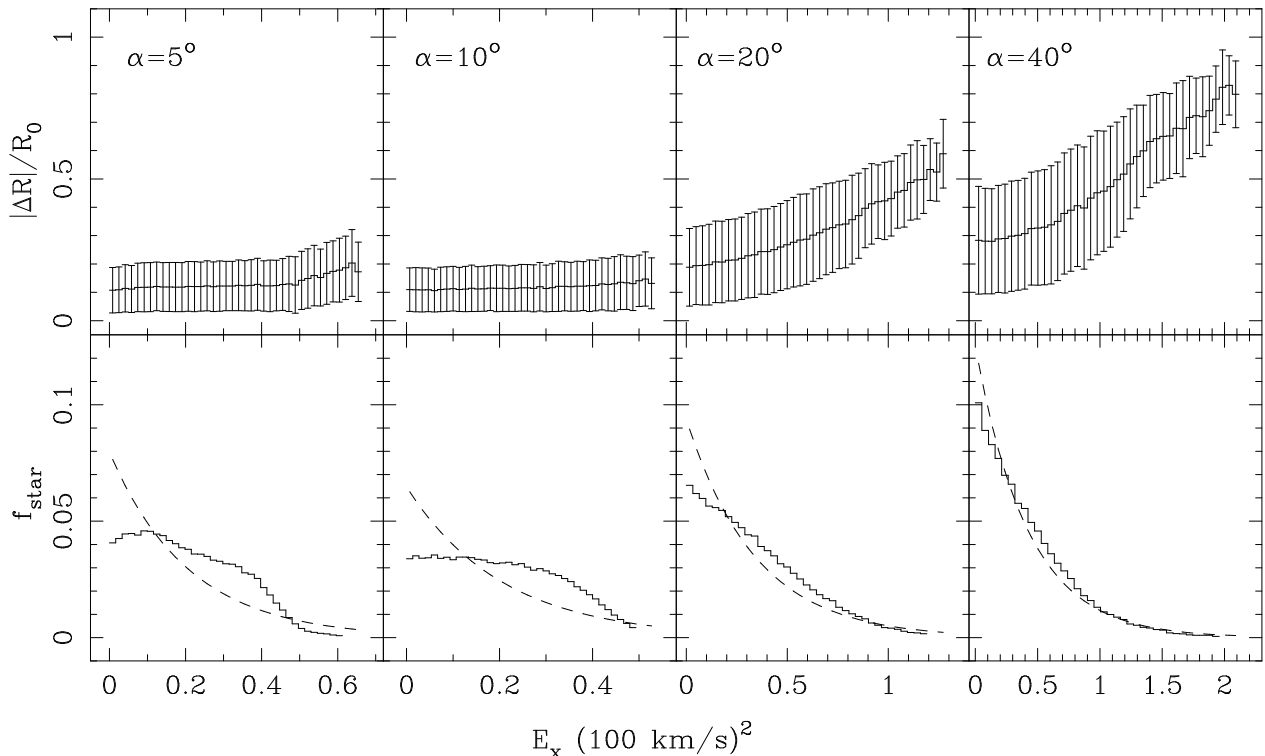}
\caption{(top) The mean of $|\Delta
R|$, the absolute value of the change in guiding-centre radius, as a function
of final epicycle energy $E_{x}=\frac{1}{2} \kappa^2 a^2$, where $a$ is the
epicycle amplitude (eq.\ \ref{eq:epicycle-energy}). The stars are initially in
circular orbits with radius $R_0$. The four panels correspond to scattering by
spiral arms with the parameters of runs \r{5},
\r{10}, \r{20} and \r{40} (Table \ref{phys-param}).  The error bars represent
the standard deviation in the distribution of $|\Delta R|$ at a given radial
energy.  (bottom) The distribution of stars as a function of epicycle
energy. Dashed lines are the Schwarzschild distribution with the same velocity
dispersion.}
\label{comp_to_binney}
\end{figure*}

We have examined the radial migration or diffusion of stars that is induced by
spiral-wave scattering, using the spiral-arm parameters for our runs \r5,
\r{10}, \r{20} and \r{40}. In each case we started 100,000 test stars in
circular orbits near the solar neighbourhood ($\pm0.2\kpc$) and followed them
for 10 Gyr. In Figure \ref{comp_to_binney} we plot the \rms\ change in
guiding-centre radius and the distribution of stars as functions of final
epicycle energy. The present in-plane epicycle energy of the Sun is only
$E_x=64(\kms)^2$, which is nearly zero on the scale of these plots. If these
simulations correctly represent the properties of the local spiral structure,
we expect that the radial migration of the Sun since its formation 4.5 Gyr ago
is in the range $0.45\pm0.25\kpc$ (pitch angle $10\degr$) to $1.4\pm0.9\kpc$
(pitch angle $20\degr$). The sheared sheet is symmetric in $\pm x$ so our
simulations cannot address the question of whether the migration is likely to
be inward or outward;
\citet{sb02} argue that migration is generally outward because of the
surface-density gradient in the disk. These estimates of the migration are
consistent with the estimate made by \citet{WFD96}, on the basis of the Sun's
metallicity and the radial metallicity gradient in the Galactic disk, that the
Sun has migrated outward by $1.9\pm0.9\kpc$ in the past 4.5 Gyr. 

\section{Discussion} \label{sec:CONCLUDE}

We have explored the hypothesis that transient spiral arms are the dominant
mechanism that drives the evolution of the velocity distribution in the solar
neighbourhood \citep{bw67,sc84,cs85}. The most thorough investigations of this
mechanism so far have been based on the Fokker-Planck equation
\citep{jb90,jen92}. Our investigation is based on direct
numerical integration of test-particle orbits in the sheared sheet
(eqs. \ref{eq:y-frame}). We impose a randomly distributed sequence of trailing
$m=2,4$ spiral waves with pitch angle $\alpha$ and a strength that is Gaussian
in time (eq. \ref{eq:phisdef}), and apply the boundary conditions that the
test particles must be on circular orbits at their formation time and in the
solar neighbourhood at the present time.

We confirm that transient spiral waves can heat the galactic disk in velocity
space. The configurations of the spiral waves in our simulations, such as 
the number ($N_s=9$--48), duration ($\Omega\sigma_s=0.5$--2.5),
\rms\ fluctuation amplitude ($\epsilon_\rms=0.16$--0.86), and pitch angle
($\alpha=5\degr$--$40\degr$) are all plausible. Spiral arms can lead to a wide
range in the observed exponent of the age-velocity dispersion relation (eq.\
\ref{eq:hist-fit}), $p=0.2$--0.76, so this exponent is not a strong
discriminator between different mechanisms of disk heating. The variations of
the vertex deviation, mean radial velocity, and axis ratios of the velocity
ellipsoid in the solar neighbourhood are consistent with observations.

The small-scale structure of the local stellar velocity distribution cannot be
interpreted in terms of axisymmetric, steady-state models.  Our simulated
distribution functions do not have the Schwarzschild form (\ref{eq:sss}) and
in addition exhibit rich small-scale structure, similar to the ``moving
groups'' and ``branches'' that observers describe in the solar neighbourhood
distribution function. This result suggests that moving groups arise primarily
from smooth star formation in a lumpy potential, in contrast to the
traditional assumption that they reflect lumpy star formation in a smooth
potential---although of course elements of both pictures must be present to
some extent.

All these findings strongly support the conclusion that transient spiral waves
are the dominant heating mechanism of the disk in the solar neighbourhood. 

Strong substructure in the local velocity distribution is most easily produced
when the Galaxy experiences a small number of short-lived but strong
(fractional surface-density amplitude $\epsilon\sim 1$) spiral transients, as
might be caused by minor mergers. A model in which spiral transients 
are continuously present at a lower amplitude could reproduce the age-velocity
dispersion relation without significant substructure.  

We also find some other results:

\begin{enumerate}

\item The age-velocity dispersion relation depends strongly on the stochastic
properties of the spiral waves, especially in cases where the heating is due
mainly to a relatively small number of spiral transients (e.g. $\sim 10$ in
runs \r{20} and \r{40}). Therefore, it is difficult, if not impossible, to
infer the details of the heating mechanism from observations of the
age-velocity dispersion relation, no matter how accurate. 

\item Spiral arms lead to radial migration of stars (\citealt{sb02} and
references therein). For the spiral-wave parameters we used, the \rms\ radial
migration of a star like the Sun is a strong function of pitch angle $\alpha$,
ranging from $0.45\pm0.25\kpc$ for $\alpha=10\degr$ to $1.4\pm0.9\kpc$
for $\alpha=20\degr$. In other words the relation between the amount of
heating of the solar neighborhood and the amount of radial migration is not
unique, a conclusion already reached by \citet{sb02} using other arguments. 

\item Stars in old moving groups did not form at a common place and time, and
therefore do not necessarily share a common age and metallicity. 

\item The traditional model, in which structure in the velocity distribution
function is the result of inhomogeneous star formation, is difficult to
reconcile with at least two observations: 1. In this model, all of the stars
in a moving group should have the same age and metallicity, while in fact the
stars in prominent moving groups show a wide range of ages. 2. It is
difficult to find a suitable astrophysical source that is small enough, cold
enough, and produces enough stars to be the progenitor of the old moving
groups.

\item We have also investigated heating by giant molecular clouds alone, and
find that this process produces much weaker substructure in the distribution
function than is observed \citep{des00}; thus giant molecular clouds are
unlikely to be the dominant heating mechanism in the solar neighbourhood. 

\end{enumerate}

Although heating by spiral waves explains most of the properties of the solar
neighbourhood velocity distribution, there is still room for other heating
mechanisms, such as the Galactic bar, giant molecular clouds, and halo
substructure or minor mergers. 

\section{Acknowledgments}

We thank James Binney, Walter Dehnen, and Jerry Sellwood for discussions and
advice, and the referee, Agris Kalnajs, for many insightful comments. This
research was supported in part by NSF Grant AST-9900316.

\end{document}